\documentclass[nofootinbib]{article}
\usepackage[total={6.5in,9in},top=1in,headsep=0.1in,headheight=1in]{geometry}
\usepackage[utf8]{inputenc}
\usepackage{hyperref}
\usepackage{amssymb}
\usepackage{graphicx}
\usepackage{lipsum}
\usepackage{hyperref}
\usepackage[maxbibnames=10,style=numeric-comp,sorting=none]{biblatex}
\usepackage{upgreek}
\hypersetup{hidelinks}
\usepackage[hybrid]{markdown}
\usepackage{color}

\newcommand{\eg}{\mbox{\emph{e.g. }}}

\newcommand{\geant}{\mbox{\texttt{Geant4}}}
\newcommand{\fluka}{\mbox{\texttt{FLUKA}}}
\newcommand{\mcnp}{\mbox{\texttt{MCNP}}}
\newcommand{\nest}{\mbox{\texttt{NEST}}}
\newcommand{\gfcmp}{\mbox{\texttt{G4CMP}}}

\newcommand\snowmass{
\begin{center}
  \rule[-0.2in]{\hsize}{0.01in}\\
  \rule{\hsize}{0.01in}\\
  \vskip 0.1in
  Submitted to the Proceedings of the US Community Study\\ 
  on the Future of Particle Physics (Snowmass 2021)\\
 
  \rule{\hsize}{0.01in}\\
  \rule[+0.2in]{\hsize}{0.01in}\\[-2em]
\end{center}
}

\usepackage[firstpage=true]{background}
\backgroundsetup{contents={\parbox{6.5in}{\snowmass}}, scale=1,placement=top,opacity=1,color=black,position={3.25in,1.2in}}

\usepackage{fancyhdr}
\fancypagestyle{plain}{%
  \fancyhf{}%
  \fancyhead[C]{}
  \fancyfoot[C]{\thepage}
}

\fancypagestyle{empty}{%
  \fancyhf{}%
  \fancyhead[C]{{\it Report of the Topical Group on Particle Dark Matter for Snowmass 2021}}
  \fancyfoot[C]{\thepage}
}
\title{Report of the Topical Group on Particle Dark Matter for Snowmass 2021}
\date{}

\usepackage{authblk}

\author[1,2]{\textbf{Conveners:} Jodi Cooley}

\author[3]{Tongyan Lin}

\author[4]{W. Hugh Lippincott}

\author[5]{Tracy R. Slatyer}

\author[6]{Tien-Tien Yu}

\affil[1]{Department of Physics, Southern Methodist University, Dallas TX 75275, USA}
\affil[2]{SNOLAB, Lively, ON P3Y 1N2, Canada}
\affil[3]{Department of Physics, University of California, San Diego, CA 92093, USA}
\affil[4]{Department of Physics, University of California, Santa Barbara, CA 93117, USA}
\affil[5]{Center for Theoretical Physics, Massachusetts Institute of Technology, Cambridge, MA 02139, USA}
\affil[6]{Department of Physics and Institute for Fundamental Science, University of Oregon, Eugene, OR 97403, USA}
\author[7]{\\ \textbf{Contributors:} Daniel S. Akerib}
\affil[7]{SLAC National Accelerator Laboratory, Menlo Park, CA 94025, USA}

\author[8]{Tsuguo Aramaki}
\affil[8]{Department of Physics, Northeastern University, Boston, MA 02115, USA}

\author[9]{Daniel Baxter}
\affil[9]{Fermi National Accelerator Laboratory, Batavia, IL 60510, USA}

\author[10]{Torsten Bringmann}
\affil[10]{Department of Physics, University of Oslo, N-0316 Oslo, Norway}

\author[11]{Ray Bunker}
\affil[11]{Pacific Northwest National Laboratory, Richland, Washington 99352, USA}

\author[12]{Daniel Carney}
\affil[12]{Physics Division, Lawrence Berkeley National Laboratory, Berkeley, CA 94720, USA}

\author[13]{Susana Cebri\'an}
\affil[13]{Centro de Astropart\'iculas y F\'isica de Altas Energ\'ias (CAPA), Universidad de Zaragoza, 50009 Zaragoza, Spain}

\author[14]{Thomas Y. Chen}
\affil[14]{Fu Foundation School of Engineering and Applied Science, Columbia University, New York, NY 10027, USA}

\author[15]{Priscilla Cushman}
\affil[15]{School of Physics and Astronomy, University of Minnesota, Minneapolis, MN 55455, USA}

\author[16]{C.E. Dahl}
\affil[16]{Department of Physics and Astronomy, Northwestern University, Evanston, Illinois 60208, USA}

\author[17]{Rouven Essig }
\affil[17]{C.N. Yang Institute for Theoretical Physics, Stony Brook University, Stony Brook, NY 11794, USA}

\author[7]{Alden Fan}

\author[18]{Richard Gaitskell}
\affil[18]{Department of Physics, Brown University, Providence, RI 02912, USA}

\author[19]{Cristano Galbiati}
\affil[19]{Physics Department, Princeton University, Princeton, NJ 08544 USA}

\author[20]{Graciela B. Gelmini}
\affil[20]{Department of Physics and Astronomy, University of California Los Angeles, Los Angeles, CA 90095, USA}

\author[21]{Graham K. Giovanetti}
\affil[21]{Physics Department, Williams College, Williamstown, MA 01267, USA}

\author[22]{Guillaume Giroux}
\affil[22]{Department of Physics, Queen’s University, Kingston, K7L 3N6, Canada}

\author[23]{Luca Grandi}
\affil[23]{Department of Physics, The University of Chicago, Chicago, IL 60637, USA}

\author[24]{J. Patrick Harding}
\affil[24]{Physics Division, Los Alamos National Laboratory, Los Alamos, NM, USA}

\author[12]{Scott Haselschwardt}

\author[9]{Lauren Hsu}

\author[25]{Shunsaku Horiuchi}
\affil[25]{Center for Neutrino Physics, Department of Physics, Virginia Tech, Blacksburg, VA 24061, USA}

\author[26]{Yonatan Kahn}
\affil[26]{Department of Physics, University of Illinois Urbana-Champaign, Urbana, IL 61801, USA}

\author[27]{Doojin Kim}
\affil[27]{Mitchell Institute for Fundamental Physics and Astronomy, Department of Physics and Astronomy, Texas A\&M University, College Station, TX 77843, USA}

\author[28]{Geon-Bo Kim}
\affil[28]{Lawrence Livermore National Laboratory, Livermore, CA 94550, USA}

\author[12]{Scott Kravitz}

\author[29]{V. A. Kudryavtsev}
\affil[29]{Department of Physics and Astronomy, University of Sheffield, Sheffield, S3 7RH, United Kingdom}

\author[7]{Noah Kurinsky}

\author[30]{Rafael F. Lang}
\affil[30]{Department of Physics and Astronomy, Purdue University, West Lafayette, IN 47907, USA}

\author[7]{Rebecca K. Leane}

\author[31]{Benjamin V. Lehmann}
\affil[31]{Department of Physics, University of California Santa Cruz, Santa Cruz, CA 95064, USA}

\author[32]{Cecilia Levy}
\affil[32]{Department of Physics, University at Albany SUNY, Albany NY 12222, USA}

\author[30]{Shengchao Li}

\author[11]{Ben Loer}

\author[12]{Aaron Manalaysay}

\author[33]{C. J. Martoff}
\affil[33]{Department of Physics, Temple University, Philadelphia, PA 19122, USA}

\author[34]{Gopolang Mohlabeng}
\affil[34]{Department of Physics and Astronomy, University of California, Irvine, CA 92697, USA}

\author[7,35,36]{M.E. Monzani}
\affil[35]{Kavli Institute for Particle Astrophysics and Cosmology, Stanford University, Stanford CA, USA}
\affil[36]{Vatican Observatory, Castel Gandolfo, V-00120, Vatican City State}

\author[37]{Alexander St\,J. Murphy}
\affil[37]{School of Physics and Astronomy, University of Edinburgh, Edinburgh, UK}

\author[38]{Russell Neilson}
\affil[38]{Department of Physics, Drexel University, Philadelphia PA 19072, USA}

\author[5]{Harry N. Nelson}

\author[39]{Ciaran A. J. O'Hare}
\affil[39]{School of Physics, The University of Sydney and ARC Centre of Excellence for Dark Matter Particle Physics, NSW 2006, Australia}

\author[40]{K.J. Palladino}
\affil[40]{Department of Physics, University of Oxford, Denys Wilkinson Building, Keble Road, Oxford, UK,  OX1 3RH}

\author[41]{Aditya Parikh}
\affil[41]{Department of Physics, Harvard University, Cambridge, MA 02138, USA}

\author[42]{Jong-Chul Park}
\affil[42]{Department of Physics and IQS, Chungnam National University, Daejeon 34134, Republic of Korea}

\author[43,44]{Kerstin Perez}
\affil[43]{Department of Physics, Columbia University, New York, NY 10027, USA}
\affil[44]{Department of Physics, Massachusetts Institute of Technology, Cambridge, MA 02139, USA}

\author[31,45]{Stefano Profumo}
\affil[45]{Santa Cruz Institute for Particle Physics, University of California, Santa Cruz, CA 95064, USA}

\author[46]{Nirmal Raj}
\affil[46]{TRIUMF, 4004 Wesbrook Mall, Vancouver BC V6T 2A3, Canada}

\author[44]{Brandon M. Roach}

\author[47]{Tarek Saab}
\affil[47]{Department of Physics, University of Florida, Gainesville, FL 32611, USA}

\author[13]{Mar\'{\i}a Luisa Sarsa}

\author[48]{Richard Schnee}
\affil[48]{South Dakota School of Mines \& Technology, Rapid City, SD 57701, USA}

\author[4]{Sally Shaw}

\author[49]{Seodong Shin}
\affil[49]{Department of Physics, Jeonbuk National University, Jeonju, Jeonbuk 54896, Republic of Korea}

\author[50]{Kuver Sinha}
\affil[50]{Department of Physics and Astronomy, University of Oklahoma, Norman, OK 73019, USA}

\author[9]{Kelly Stifter}

\author[12]{Aritoki Suzuki}

\author[32]{M.~Szydagis}

\author[34]{Tim M.P. Tait}

\author[51,52]{Volodymyr Takhistov}
\affil[51]{Kavli Institute for the Physics and Mathematics of the Universe (WPI), UTIAS}
\affil[52]{University of Tokyo, Kashiwa, Chiba 277-8583, Japan}

\author[34]{Yu-Dai Tsai}

\author[53]{S.~E.~Vahsen}
\affil[53]{Department of Physics and Astronomy, University of Hawai`i, Honolulu, HI 96822, USA}

\author[20]{Edoardo Vitagliano}

\author[53]{Philip von Doetinchem}

\author[54]{Gensheng Wang}
\affil[54]{High Energy Physics Division, Argonne National Laboratory, Lemont, IL 60439, USA}

\author[55]{Shawn Westerdale}
\affil[55]{Department of Physics and Astronomy, University of California, Riverside, California 92521, USA}

\author[31,44]{David A.\ Williams}

\author[56]{Xin Xiang}
\affil[56]{Brookhaven National Laboratory, Upton, NY, USA}

\author[3]{Liang Yang}
\begin{document}

\maketitle
\section{Executive Summary}

One of the most important scientific goals of the next decade is to reveal the nature of dark matter (DM). 
To accomplish this goal, we must {\textbf{delve deep}}, to cover high priority targets including weakly-interacting massive particles (WIMPs), and {\textbf{search wide}}, to explore as much motivated DM parameter space as possible. A diverse, continuous portfolio of experiments at large, medium, and small scales that includes both direct and indirect detection techniques maximizes the probability of discovering particle DM.  Detailed calibrations and modeling of signal and background processes are required to make a convincing discovery. In the event that a candidate particle is found through different means, for example at a particle collider, the program described in this report is also essential to show that it is consistent with the actual cosmological DM. The US has a leading role in both direct and indirect detection dark matter experiments -- to maintain this leading role, it is imperative to continue funding major experiments and support a robust R\&D program.

\subsection*{\textit{Dark Matter exists}}
We have strong astronomical and cosmological evidence across many scales, ranging from velocities of stars in ultra-faint galaxies to the Cosmic Microwave Background, indicating that $>$80\% of the matter in our Universe consists of non-baryonic matter known as ``Dark Matter".   The steady collection of this evidence over the last eight decades suggests that DM consists of one or more new particles not contained within the Standard Model (SM) of Particle Physics. From these observations, we can set some basic requirements on DM. The dominant part of DM must be stable with a lifetime much longer than the age of the Universe, it must form cosmological structures consistent with observations, and it must be produced in the early Universe. From these requirements, we can infer some of the general properties of DM. The formation of consistent cosmological structures sets a limit on the strength of DM interactions, both with the SM and with itself. DM production in the early Universe may occur through a wide range of channels, but many of these require additional non-gravitational interactions in some form.

\subsection*{\textit{Particle dark matter is theoretically well-motivated}}  

There is a wealth of scenarios for DM that are consistent with these observations and requirements. 
The range of experimentally allowed DM candidates spans masses from $m_\chi\sim 10^{-21}$ eV, which is the lightest mass that is still consistent with galactic structure, to that of macroscopic objects which can be several hundreds of solar masses. 
This report focuses on DM candidates with masses in the eV to 100 TeV range, with a subset of candidates extending out to the Planck scale ($m_\chi\sim M_{\rm Pl}=10^{19}$ GeV).
In this mass range, DM candidates are best described by a particle picture and we denote these candidates as {\bf particle dark matter}. 
Theoretical paradigms for particle DM that remain highly-motivated include those of WIMPs, asymmetric DM, and hidden sectors which span the entirety of the mass range.
For a given cosmological history of the DM, these paradigms can be quite predictive, giving rise to specific DM mass scales and interactions. 

\subsection*{\textit{``Delve Deep and Search Wide": A diverse portfolio maximizes the probability of discovery}}

To understand the {\it particle} nature of DM, we must employ multiple experimental techniques spanning the entire range of the High-Energy Physics (HEP) program, including both direct and indirect detection techniques. Moderate- and large-scale experiments will allow us to delve deep into high priority target scenarios such as WIMPs, whereas a well-chosen ensemble of small-scale experiments provides versatility and the ability to test an expanded range of models. R\&D towards improved detector technologies are fundamental and crucial to a comprehensive exploration of the dark sector. Direct detection experiments, particularly the next generation of WIMP searches, will require continued investment in underground facilities.

Importantly, both direct and indirect detection, and other Cosmic Frontier probes, are the only way to connect an observed DM candidate with its astrophysical environment. Indirect detection probes time and distance scales that dwarf those probed by any terrestrial detector, allowing for unique sensitivity to certain properties of DM. For example, if the DM is not absolutely stable, its decay lifetime is longer than the age of the universe and so is likely to be measurable only by indirect methods. Direct detection experiments are adaptable, and can respond readily to signals or hints or the lack thereof to mitigate systematic backgrounds or redirect to new parameter space. Furthermore, direct detection experiments can be, and often are, adjusted, upgraded, fixed, or moved to confirm or disprove any observation of DM that may arise.

Crucially, DM is a major science driver across {\it all} of the experimental frontiers of particle physics: Cosmic, Energy, Rare and Precision, and Neutrino, as well as the cross-cutting Accelerator, Instrumentation, Underground, and Theory Frontiers. As the search for DM transcends frontier boundaries, a unified and complementary strategy is needed to maximize discovery, as described in~\cite{Complementarity}.
\subsection*{\textit{Understanding how signals and backgrounds manifest in an experiment is essential to making a detection}}
Precise and accurate modeling of both signal and background, informed by calibration and propagated via simulation, is essential to confirm and validate any current or future potential DM signal.
For indirect detection experiments, it is necessary to have an improved understanding of a range of astrophysical backgrounds and processes, including as one example cosmic ray production and propagation. For direct detection experiments, future advances will require an improved understanding of threshold effects, radiogenic backgrounds, dark noise, the ways DM interacts in various materials, and the optical and atomic properties of materials.

\section{Introduction}
\label{sec:intro}
This report is primarily based on community input on big questions in the search for particle DM, summarized in eight solicited Snowmass white papers. 
These papers focused, respectively, on the following topics:
\begin{itemize}
    \item {\bf WP1} -- Dark matter direct detection to the neutrino fog~\cite{CF1WP1}
    \item {\bf WP2} -- The landscape of low-threshold dark matter direct detection in the next decade~\cite{CF1WP2} 
    \item {\bf WP3} -- Calibrations and backgrounds for dark matter direct detection~\cite{CF1WP3}
    \item {\bf WP4} -- Modeling, statistics, simulations, and computing needs for direct dark matter detection~\cite{CF1WP4}
    \item {\bf WP5} -- The landscape of cosmic-ray and high-energy photon probes of particle dark matter~\cite{CF1WP5}
    \item {\bf WP6} -- Puzzling excesses in dark matter searches and how to resolve them~\cite{CF1WP6}
    \item {\bf WP7} -- Synergies between dark matter searches and multiwavelength/multimessenger astrophysics~\cite{CF1WP7}
    \item {\bf WP8} -- Ultraheavy particle dark matter~\cite{CF1WP8}
\end{itemize}
Over twenty other white papers were submitted to CF1 (in many cases cross-listed with other working groups) and can be found via the Snowmass website: \url{https://snowmass21.org/submissions/cf} -- we also reference them where appropriate below. 

\subsubsection*{Dark matter is a fundamental mystery of high energy physics}

We have strong astronomical and cosmological evidence across many scales, ranging from the velocities of stars in ultra-faint galaxies to the Cosmic Microwave Background, which indicates that $>80$\% of the matter in our Universe consists of non-baryonic matter known as ``dark matter"~\cite{Bertone:2016nfn}. The steady collection of this evidence over the last century strongly indicates that DM is a new particle not contained within the Standard Model of Particle Physics, that the dominant part of DM must be stable with a lifetime much longer than the age of the Universe, and that the DM was produced in the early Universe. 
This last feature may provide a clue to non-gravitational interactions of the DM. 
Furthermore, the DM must form cosmological structures consistent with observations, which in turn sets a limit on the strength of the DM interactions with the SM and with itself.

Understanding the nature of DM is a top priority within the HEP community. In 2014, the Particle Physics Project Prioritization Panel (P5) identified the search for DM as one of the five priority science drivers for the HEP Program~\cite{P5Report:2014pwa}. ``Dark matter'' was the most common phrase among the Letters of Interest submitted by the entire HEP community in the run up to Snowmass 2022~\cite{WattsWebsite}. Dark matter is a fundamental mystery of high energy physics.

\subsubsection*{There is a wide range of viable particle DM candidates}
There are many theoretically motivated DM candidates capable of explaining the astrophysical evidence for DM, which span mass ranges from the lightest masses that are still consistent with galactic structure ($m_\chi\sim 10^{-21}$ eV) to macroscopic objects that can be several hundreds of solar masses~\cite{Asadi:2022njl}. This report focuses on particle DM candidates with masses from the eV range to the Planck scale ($m_\chi\sim M_{\rm Pl}=10^{19}$ GeV).
In this mass range, DM candidates are best described by a particle picture and we denote these candidates as {\bf particle DM}.
\kern-.20em\footnote{For DM masses above an eV, the occupation number of particles within a volume of radius equal to the de Broglie wavelength is around unity or smaller ($n_{\chi}\lambda_{\rm dB}^3\ll 1$). For mass-scales well below an eV, the occupation number is much higher and DM is better described by a wave picture ($n_{\chi}\lambda_{\rm dB}^3\gg 1$). This latter category of DM candidates, which contains the QCD axion, is covered by the topical working group on Wave-Like Dark Matter (CF2).}

The eV-$M_{\rm Pl}$ range of particle DM candidates can be further sub-divided into three categories:
\begin{itemize}
    \item 1 eV$\lesssim m_\chi\lesssim 1$ GeV: ``Light" DM 
    \item 1 GeV$\lesssim m_\chi\lesssim 100$ TeV: ``Heavy" DM 
    \item $m_\chi\gtrsim 100$ TeV: ``Ultra-Heavy" DM (UHDM)
\end{itemize}

The most well-known DM candidates in this mass range are the weakly-interacting massive particles (WIMPs) which have masses between 1 GeV to 100 TeV, interact weakly with the SM, and fall into the ``heavy" DM category~\cite{Bertone:2016nfn}. Such candidates are strongly motivated in particle physics, frequently appear in supersymmetric (SUSY) models designed to address the gauge hierarchy problem
\hspace{-0.1cm}\footnote{The large discrepancy between the electroweak scales of ${\cal O}$(100 GeV) and $M_{\rm Pl}$ is known as the gauge hierarchy problem. The resolution of this problem remains one of the outstanding issues of particle physics.}, and are a high-priority target for particle DM searches.
WIMPs are also appealing on cosmological grounds as they lead to the correct relic density of DM through either mechanisms of thermal freeze-out (``WIMP miracle")~\cite{Jungman:1995df,Bertone:2004pz} or through a connection of the DM
 asymmetry to the baryon-antibaryon asymmetry (``asymmetric DM")~\cite{Kaplan:2009ag}. The latter mechanism naturally accommodates DM candidates with masses around a few GeV but this range can be extended to both lower and higher masses. Although not yet observed despite the advances in sensitivity made by accelerator and cosmic DM searches over the past decade, WIMPs remain a well-motivated DM candidate providing a rich suite of accessible targets for next generation experiments. 
 
There are many classes of models for UHDM candidates, including WIMPzillas, strangelets, and Q-balls~\cite{Guepin:2021qai,CF1WP8, Asadi:2022njl}. Non-thermal UHDM candidates, known as WIMPzillas, are a generic phenomenon of inflationary cosmology~\cite{Chung:1998zb,Kolb:1998ki}, whereas $Q$-balls (\eg~\cite{hepph0303065}) provide a candidate that explains both baryogenesis and DM, with a potential connection to inflation. Nuggets of quarks formed in first-order phase transitions result in strangelets that are viable DM candidates~\cite{Carlson:2015daa}. 
 
In the last decade, the DM community has realized that DM may be just one particle of many that comprise a ``hidden sector" that contains its own matter and force particles. Importantly, a hidden sector with typical mass scales of a few tens of GeV or lower must be only feebly coupled to the SM. One can systematically write down the renormalizable couplings between a hidden sector and the SM, which define {\bf portals} that lead to distinct and predictive interactions between DM and the SM~\cite{Hewett:2012ns,Essig:2013lka,Alexander:2016aln,Battaglieri:2017aum,Jaeckel:2010ni,Essig:2022yzw,Zaazoua:2021xls,Aboubrahim:2021dei}.  As such, models of hidden sector DM contain rich phenomenology to search for the DM, the mediator, and interactions within a hidden sector itself. DM candidates that live in a hidden sector can span the entire range of particle DM masses but are especially of interest for candidates of ``light" DM.
The observation of non-zero neutrino masses motivates the existence of right-handed or sterile neutrinos. For certain ranges of masses and interaction strengths, these neutrinos are also candidates for DM with masses ranging from the Tremaine-Gunn bound of a few hundred eV to tens of keV~\cite{Dodelson:1993je,Abazajian:2012ys,Abazajian:2017tcc}.  These BSM neutrinos provide another class of motivated and predictive ``light" DM candidates.
\subsubsection*{Diversity of project scales maximizes our probability of discovering DM}
As we see, there are several theoretically motivated avenues for DM candidates that lead to strikingly different observational signatures.
The strong observational evidence which supports the existence of DM is all a result of the gravitational influence of DM on visible matter. To understand the {\it particle} nature of DM, we must also understand its non-gravitational interactions, which requires probes of all potential DM interactions: with photons, electrons, neutrinos, and quarks, or with other unknown particles that live within a hidden sector.  

As such, uncovering the particle nature of DM requires a multi-pronged search strategy to thoroughly study all its potential non-gravitational interactions including: the interaction of DM with targets in a detector ({\bf direct detection}); through the detection of SM particles that result from decays or annihilations of DM ({\bf indirect detection}); through the production of DM in colliders or accelerators; and through the effects of DM-SM interactions on astrophysical systems. Within direct detection, experimental signatures include {\it scintillation photons, ionization electrons}, and {\it phonons}. For indirect detection, the experimental signatures include {\it photons, cosmic-rays,} and {\it neutrinos} at energy scales from keV-ZeV. In addition to these two classes of DM probes, there are cosmic and accelerator/collider probes. Discussion of these additional probes can be found in the reports of topical groups working on cosmic probes of DM (CF3) and fundamental physics more broadly (CF7), DM at colliders (EF10), and dark sector studies at high intensities (RF6).

The P5 recommendations of 2014 included increased funding of the then-proposed generation of direct detection experiments (sometimes called ``Generation 2
" or ``G2''), support for one or more next generation experiments, maintaining a program at smaller scale to support new ideas and developments, and possible investment in the Cherenkov Telescope Array (CTA) experiment. The G2 experiments are now in operation or near the end of construction, but unfortunately there has been no movement in the US towards a next generation program. In 2018, the Basic Research Needs Dark Matter New Initiatives program (DMNI) ~\cite{BRNreport} led to the development of several promising direct detection ideas; this process was very valuable to the community and provides a useful model for supporting smaller projects of all kinds going forward. Unfortunately, the current DMNI experiments also await the next stage of funding to proceed to the project phase.

Today, the field of particle DM is still growing rapidly, generating promising ideas to probe DM in all of its possible variations. A diverse, continuous portfolio of experiments that includes both direct and indirect detection techniques at multiple scales maximizes the probability of discovering particle DM.  Detailed calibrations and modeling of signal and background processes, as well as independent experimental confirmation of a putative signal, are required to make a convincing discovery. The US has a leading role in both direct and indirect detection DM experiments, and it is imperative to support next generation DM programs and maintain a robust R\&D program to develop technologies that enable this pursuit.

The remainder of this report will summarize the status and prospects for observing particle DM for the two categories of indirect and direct detection.
\section{Indirect Detection}
\subsection{Introduction}

Indirect probes of particle DM are generally those that search for the signatures of DM interactions throughout the cosmos, in contrast to direct probes which search for DM interactions within a detector volume. The primary program of {\it indirect detection} is to search for ``cosmic messengers'' produced directly by DM interactions, in particular via annihilation or decay of the DM. Indirect detection searches have traditionally relied on photon and cosmic ray messengers due to their relative ease of detection. However, the rise of neutrino observatories as well as the detection of gravitational waves have augmented the family of messengers to include neutrinos and gravitational waves, \eg \cite{Berti:2022wzk}. There are also powerful cosmic probes of DM using ``natural laboratories'', where astrophysical objects are affected by the presence and/or interactions of DM, which can have important commonalities with ``cosmic messenger'' searches; we discuss some aspects of these searches briefly, while ceding a more in-depth discussion to papers from the dedicated topical group on Cosmic Probes of Dark Matter, \eg\cite{SnowMassWPxtremenviron}. 

Indirect detection probes time and distance scales that dwarf those probed by any terrestrial detector, allowing for unique sensitivity to certain properties of DM; for example, if the DM is not absolutely stable, its decay lifetime is longer than the age of the universe and so is likely to be measurable only by indirect methods. 

Typically annihilation and decay of DM produce signals with characteristic energy scales set by the DM mass, although there are also frequently secondary signals at much lower energies (\eg arising from synchrotron and bremsstrahlung from electrons and positrons~\cite{Profumo:2010ya}, or pair production from very-high-energy photons~\cite{Blanco:2018bbf,Heiter:2017cev,AlvesBatista:2016vpy,Murase:2015gea,Blanco:2018esa}); this allows indirect searches to probe DM at much higher masses than their nominal energy scales. There are existing (and proposed new) technologies for measuring cosmic messengers over an enormous range of energy scales, from sub-eV to ZeV, providing sensitivity to a correspondingly wide range of DM masses. Especially for higher DM masses, where abundant energy is available to produce the full range of Standard Model particles, indirect signals are generically expected to be multi-messenger and multi-scale~\cite{CF1WP8}. As a result, a broad program of multi-messenger and multi-wavelength probes both maximizes sensitivity and allows powerful internal consistency checks on a putative signal, which are often quite model-independent.

At the same time, indirect searches frequently encounter large systematic uncertainties and backgrounds surrounding the DM and its environment. A confirmed detection of a DM signal will likely require robust characterization of emission unrelated to DM, so that the signal of interest may shine through. Multiwavelength and multimessenger observations and analysis offer the promise of helping identify, quantify, and remove these backgrounds to DM searches. In particular, there are currently a number of detected excesses above currently-understood backgrounds; understanding the origin of these excesses could provide important insights into backgrounds or even reveal a DM signal. 

\subsection{Science targets for indirect detection}

As discussed in Sec.~\ref{sec:intro}, it is convenient to subdivide particle DM candidates by their mass into ``light'' (sub-GeV), ``heavy'' or ``electroweak scale'' (GeV--100 TeV), and ``ultra-heavy'' ($\gg 100$ TeV). WIMPs, falling in the ``heavy'' range, are perhaps the longest-studied candidate, and can be probed synergistically by indirect, direct, and collider searches. In this mass range, one widely-used benchmark for indirect detection is the thermal freeze-out mechanism, where the annihilation rate of DM in the early universe controls its late-time abundance. 

Under assumptions of a standard cosmological history, this classic scenario requires a nearly mass-independent annihilation cross section of $2-5\times 10^{-26}$ cm$^3$/s~\cite{Gondolo:1990dk,Steigman:2012nb} for self-conjugate dark matter annihilating through a 2-body process with a cross section that is close to velocity-independent in the non-relativistic limit.\footnote{The required cross-section is approximately doubled if the dark matter and anti-matter are separate particles with equal abundances.} If the cross section does not evolve with time, then this mechanism has no free parameters and directly predicts an annihilation cross section that can be observed today. 

In principle, this is a simple and viable way of generating the observed relic abundance for DM masses as light as 1 MeV or as heavy as 100 TeV, without modifying the cosmological history; at lower masses, some additional ingredients are typically needed to avoid disruption of Big Bang nucleosynthesis, which occurs at MeV-scale temperatures, \eg~\cite{Sabti:2019mhn}. At masses above 100 TeV, unitarity sets largely model-independent upper limits on the annihilation cross section during freezeout, leading to a prediction that thermal relic DM should overclose the universe, \eg~\cite{Griest:1989wd, Smirnov:2019ngs}. Current constraints typically exclude this thermal relic cross section, for annihilation into any Standard Model final states except neutrinos, for DM masses below 10-100 GeV, \eg~\cite{Leane:2018kjk}.

There are many variations on this scenario (a more in-depth discussion of the theoretical possibilities may be found in \cite{Asadi:2022njl}). A simple modification is that the DM may be asymmetric, with less anti-DM than DM; in this case, or more generally if the annihilation requires the presence of a partner particle that is depleted in the late universe, that asymmetry may cut off annihilation at late times and strongly suppress indirect signals. The cross section may be suppressed at low velocities, or may require more than two particles in the initial state, both of which suppress indirect detection signals compared to the freeze-out epoch (characterized by much higher density and temperature). These modifications open up thermal relic parameter space in the ``light DM'' mass range. At higher masses, modifications to the assumed cosmological history can greatly modify the permissible parameter space, extending the upper mass bound to $10^{10}$ GeV or higher; examples of such modifications include an epoch of early matter domination, a phase transition involving the DM, and/or dilution of the DM by the decays of other particles, \eg~\cite{Gelmini:2006pq,Gelmini:2006pw,,Hamdan:2017psw,Howard:2021ohe,Dienes:2021woi,Berger:2020maa,Bramante:2017obj,Cirelli:2018iax,Bhatia:2020itt,Berlin:2016gtr,Berlin:2016vnh,Heurtier:2019eou,Davoudiasl:2019xeb}. Furthermore, independent of its mass scale, the DM need never achieve thermal equilibrium with the Standard Model -- thermal freeze-out is a classic benchmark with broad applicability, but far from the only possibility.

In the ``light'' sub-GeV mass range, another well-motivated target is the keV-scale sterile neutrino. Constraints from production in the early universe~\cite{Venumadhav:2015pla, Laine:2008pg}, structure formation~\cite{Cherry:2017dwu}, and X-ray telescopes~\cite{Ruchayskiy:2015onc, Malyshev:2014xqa, Tamura:2014mta, Perez:2016tcq,Ng:2019a,Roach:2019ctw,Ng:2015gfa, Boyarsky:2007ge} have narrowed the viable parameter space in the simplest models to a small region~\cite{CF1WP5}.

Beyond these benchmarks, non-thermal production scenarios include freeze-in, out-of-equilibrium decay of a heavy field, phase transitions, gravitational particle production, and decay of black holes to Planck-scale relics (see~\cite{CF1WP5,CF1WP8,Asadi:2022njl} for a more in-depth discussion of these scenarios and further references). Taken together, these scenarios can accommodate particle DM candidates across the full mass range we consider, and can allow for striking indirect signatures; while annihilation signatures are suppressed at high masses, there can still be detectable multimessenger and multiwavelength signals from decays. At mass scales below those we consider, in the wave-like regime, observations of cosmic messengers and natural laboratories can also place powerful bounds on interconversions between visible particles and very light non-thermal DM; we will not discuss such bounds further in this report, but mention them as an example of synergy.

\subsection{The path toward dark matter discovery with indirect detection}

\subsubsection{Improving discovery potential for particle dark matter across its full mass range}

Indirect searches for DM span a vast range of target regions, energy scales, and messengers. Observations of galaxy clusters, dwarf galaxies, the Andromeda galaxy, the Magellanic clouds, the Sun,  diffuse emission and possible DM clumps in the Milky Way, the extragalactic background light, and the Milky Way Galactic Center, have all been used to set limits on the interactions of DM (see~\cite{CF1WP5,CF1WP6} for further discussion and references), and future instruments will allow us to improve the sensitivity of these probes. At low energies, the strongest current limits from classic indirect detection typically arise from cosmic-ray and gamma-ray searches, but neutrino bounds can take over for decaying DM at masses above 10 TeV~\cite{CF1WP8,IceCube:2018tkk,Arguelles:2019boy}. 

Celestial bodies, as natural laboratories, may in some circumstances play an important role in mediating the signals of DM interactions -- \eg via annihilation of DM captured in the Sun, by the heating of neutron stars from DM scattering, or by accumulation of DM leading to low-mass black hole formation. Some such signals will be visible to standard indirect searches, but others will require dedicated analyses -- \eg recent or proposed  infrared telescopes such as the James Webb Space Telescope (JWST), the Thirty Meter Telescope (TMT) and the Extremely Large Telescope (ELT) can search for cold neutron stars~\cite{NSvIR:Baryakhtar:DKHNS}, and gravitational wave experiments could detect or exclude a black hole population formed by dark-matter-induced collapse~\cite{Dasgupta:2020mqg}.

{\it Given the wide range of plausible DM scenarios, it is important to continue a broad program of indirect searches with sensitivity to different energy scales and cosmic messengers.} Beyond individual observations, combining and cross-correlating searches can improve sensitivity and our ability to reject complex backgrounds. In addition, we strongly encourage experimental collaborations to display the full extent of their inferred limits along the DM mass axis;
as discussed above, secondary particle production means that measurements at a given energy scale can often be sensitive to much higher DM masses, and often these limits are artificially cut off as part of the analysis.

Within this broad framework, if we focus on cosmic messenger searches with gamma rays and charged cosmic rays (with neutrino searches being discussed in Neutrino Frontier reports, \eg \cite{Berger:2022cab}), specific directions being pursued by proposed instruments include:
\begin{itemize}
    \item Testing the thermal relic cross section across its full mass range (in the standard cosmological history), from MeV -- 100 TeV.
    \item Improving sensitivity to MeV-GeV gamma-rays, where there is currently a sensitivity gap compared to lower- or higher-energy photons, and the potential to explore wide swaths of currently unconstrained parameter space.
    \item Improving sensitivity in the X-ray band, especially to monochromatic line signals via improved energy resolution; near-future experiments may achieve sufficient resolution to resolve Doppler broadening from the velocity of DM in the Milky Way halo.
    \item Developing methods that can detect low-energy (non-relativistic) antimatter cosmic rays. A confirmed detection of low-energy antideuterons or antihelium would be transformative, as these channels are thought to be essentially background-free. Future experiments for antinuclei searches are needed to measure down to the actual astrophysical spectrum due to primary CR interactions with the ISM.
    \item Improving the characterization of key backgrounds and systematic uncertainties in a range of searches, both generally and as a step to resolving the origins of various currently-unexplained excesses.
\end{itemize}

\subsubsection{Enabling discovery with complementary measurements and modeling}
\label{sec:indirect-complementary}

For most possible indirect signals of DM interactions, an accurate understanding of relevant astrophysical processes and backgrounds is {\it essential} for reliably translating observations into constraints on -- or a discovery of -- DM physics. There are only a handful of plausible signals that are believed to be effectively background-free; examples include gamma-ray spectral lines and production of low-energy antideuterons. Exploring these channels is necessary, but other searches may have higher sensitivity if backgrounds can be controlled. Table~\ref{tab:DMsignals_vs_BackgroundUncertainties} outlines some categories of backgrounds for the specific case of gamma-ray searches. 

Consequently, there is an extensive class of measurements which are important for enabling future indirect probes of DM physics, even if they do not involve directly measuring cosmic messengers from DM interactions. We describe some of these below.

  \begin{table}[ht!]
    \centering
    \begin{tabular}{|c|c|c|c|}\hline\hline
         & Galactic diffuse & Unresolved  & Instrumental \\
         & emission &  Sources  & Systematics \\\hline         
MeV DM              & $\bullet$ & $\bullet$ & \\\hline         
GeV DM @GC          & $\bullet$ & $\bullet$ & \\\hline
GeV DM @M31         & $\bullet$ & $\bullet$ & \\\hline
GeV DM @dSphs       &           &           & $\bullet$ \\\hline
GeV DM @clusters    &           & $\bullet$ & \\\hline
Heavy DM            &           &  $\bullet$         & $\bullet$ \\\hline
    \end{tabular}
    \caption{Summary of DM searches with gamma rays and the main categories of backgrounds. Bullet points illustrate the main relevant backgrounds; other backgrounds can still be relevant under specific conditions. Here, heavy DM refers to DM above the electroweak scale. Reproduced from Ref.~\cite{CF1WP5}.}
    \label{tab:DMsignals_vs_BackgroundUncertainties}
\end{table}

\subsubsection*{Cosmic ray production, composition, and propagation}
{\it This section draws extensively on discussions in WP5, WP6, and WP7~\cite{CF1WP5,CF1WP6,CF1WP7}.}

Charged cosmic rays can arise either as cosmic messengers of DM physics or through a myriad of energetic processes in our Galaxy and beyond. Understanding how these particles are produced and then propagate from their sources to Earth is a key ingredient in both signal and background modeling for charged-particle searches. Furthermore, because the cosmic rays can produce photons and neutrinos as they interact with the interstellar gas and radiation, their behavior is also a key input for searches in these channels. In cases where the DM itself, or a particle produced by its decays, carries a tiny electric charge, magnetic fields could modify the distribution of DM and/or its signals. It is challenging to measure the magnetic field in dark structures (such as  dwarf spheroidal galaxies or the external regions of spiral galaxies and clusters), but great progress is expected in the coming years due to next-generation radio telescopes, such as the Square Kilometer Array (SKA)~\cite{SKAMagnetismScienceWorkingGroup:2020xim}.

Ongoing cosmic-ray observations of multiple species, including unstable species, can help constrain propagation, \eg~\cite{1990acr..book.....B,2007ARNPS..57..285S,2017ApJ...840..115B,Reinert:2017aga,2018ApJ...854...94B,2018ApJ...858...61B,Boudaud:2019efq,2019arXiv191103108B}. Improvements in modeling and simulation of the sources of cosmic rays, with careful study of uncertainties related both to production and propagation, would be valuable in this regard. Another direction for improvement is the full understanding of the experimental uncertainties in cosmic-ray measurements and their correlations, particularly from the AMS-02 experiment, which may require collaboration between the experimental and theoretical communities.

For positrons, in particular, recent observations of TeV gamma-ray halos around pulsars suggest that these are a major source of high-energy positrons~\cite{Abeysekara:2017science,LHAASO:2021crt}. Future observations of these halos from ground-based gamma-ray experiments such as the High Altitude Water Cherenkov Experiment (HAWC), the Large High Altitude Air Shower Observatory (LHAASO) and CTA will have the capacity to probe the injection spectrum of these positrons and the size of the low diffusion bubble present around sources. More generally, multiwavelength observations of the Galaxy in photons will continue to help constrain cosmic-ray populations via their secondary emission, and to search for the sources of cosmic rays.

The production cross sections for photons and antiparticles in cosmic ray collisions with the gas still have large uncertainties in many cases. In particular, very few direct measurements exist for antiproton production from protons interacting with heavier nuclei, and there are substantial uncertainties in the cross sections for positron production in the GeV-TeV range, via $ p + p \rightarrow  e^+ + X$ and $ p + \mathrm{He} \rightarrow  e^+ + X$. 

These uncertainties both impact predictions for backgrounds, and potentially bias the inference of cosmic ray propagation parameters from measurements of the cosmic-ray spectra at Earth. Furthermore, to the extent that the models entering into these cross sections are also used to predict antiparticle signatures from DM annihilation or decay, data that allows refinement of the models will directly impact predictions of DM signals. In particular, the yield of antideuterons and antihelium depends on the details of hadronization and coalescence once quarks are produced, either in purely Standard Model collisions or in hypothetical DM interactions.

This is an area where considerable improvement is possible in the near future. Measurements from fixed-target accelerator experiments, covering a range of
energies, with the capacity to detect antiproton, antideuteron and antihelium nuclei, will be of great benefit to achieve reduced cross-sectional uncertainties. 
Measurements of relevant cross sections have been performed and/or proposed at both current experiments such as NA61/SHINE~\cite{Aduszkiewicz:2017sei}, ALICE~\cite{ALICE:2020zhb} and LHCb~\cite{Aaij:2018svt}, and at AMBER~\cite{Denisov:2018unj} (the next-generation successor of the CERN SPS COMPASS experiment); details may be found in~\cite{CF1WP5}.

\subsubsection*{The Galactic diffuse photon emission}

{\it This section draws extensively on discussions in WP6 and WP7~\cite{CF1WP6,CF1WP7}.}

As Earth is embedded within the Milky Way Galaxy, essentially all photon-based indirect DM searches require some model for the Galactic foregrounds. These foregrounds can depend on the distribution of gas, cosmic rays, and radiation throughout the Milky Way, the turbulent and regular components of the Galactic magnetic field, and large-scale structures such as the Fermi Bubbles. 

To quote from~\cite{CF1WP7}, {\it ``the single most important challenge in generating accurate Galactic diffuse emission models remains obtaining accurate distributions of input ingredients, e.g., cosmic-ray sources, magnetic fields, and target gas densities''}. Refs.~\cite{CF1WP6, CF1WP7} discuss a number of avenues to improve the current situation, including:
\begin{itemize} 
\item Use of multiwavelength and multimessenger data, such as new maps of the interstellar dust, new constraints on the interstellar radiation field as a function of frequency, and improved measurements of the magnetic field from upcoming radio telescopes,
\item Development of improved tracers for components of the interstellar gas, particularly molecular hydrogen,
\item Improved modeling techniques, in particular avoiding unnecessary simplifying assumptions, accounting for non-steady-state behavior, incorporating insights from simulations, and performing careful error propagation to allow for marginalization over systematic uncertainties.
\end{itemize}

In addition to improving the input ingredients, sideband analyses, such as studies of MeV gamma-rays in cases where the dominant signal is expected in the GeV band, can be powerful in informing data-driven models of the backgrounds. In particular, better modeling the MeV-band diffuse emission in the inner Galaxy could test scenarios for excess emission both at higher energies (the ``GeV excess'') and lower energies (excess 511 keV emission).

In modeling the diffuse emission, it will likely be necessary to go beyond simplifying assumptions of cylindrical symmetry and conduct anisotropic, three-dimensional modeling of the diffuse emission. Promising initial work~\cite{Porter_2017,Johannesson:2018bit} remains impeded by computational challenges, and modeling and computational strides are urgently required to reduce the systematic Galactic diffuse emission uncertainties.

\subsubsection*{The dark matter density distribution}

All indirect probes of DM interactions depend on some model of the density for DM in the target region. There are substantial theoretical and observational uncertainties in these density profiles, even in the case where DM is assumed to be collisionless; these uncertainties are amplified if the DM has substantial interactions with DM or other particles.

Ref.~\cite{CF1WP7} discusses a number of efforts underway on the theory/analysis side to improve reconstruction of DM density profiles from stellar data, minimizing unwarranted simplifying assumptions. Correctly modeling baryonic physics and the associated uncertainties is also a major ongoing challenge for cutting-edge hydrodynamical simulations.

On the observational side, searches for new dwarf galaxies with the Vera Rubin Observatory have the potential to identify promising new targets for indirect detection, and Gaia stellar data can be used to constrain the DM distribution of the Milky Way (albeit with large uncertainties especially toward the Galactic Center)~\cite{CF1WP6,2020gaia}.

\subsubsection*{Other complementary measurements and analyses}

Measurements of the intergalactic photon background at a range of energies can be sensitive to DM interactions, but the dominant flux is astrophysical; measurements that constrain the properties of other extragalactic contributions to this background light allow for improved DM searches. Similarly, a better understanding of the sources of diffuse neutrino flux could strengthen neutrino constraints on very heavy DM, and a better understanding of the composition of ultra-high-energy cosmic rays could shed light on their origin and any possible new physics contributions. Particles produced in collisions of cosmic rays with the atmosphere also provide a search channel for a range of DM models. At lower energies, neutrinos serve as a background for {\it direct} detection; paleodetectors, which are discussed in Sec.~\ref{sec:DDsignatures}, could potentially probe the neutrino flux as a function of time.

Pulsars generate both cosmic rays and gamma rays and can have energy distributions that resemble those expected from annihilation/decay processes, thus providing challenging backgrounds for a range of DM indirect searches. Searches for pulsars at radio and X-ray wavelengths have a long history, and upcoming telescopes will improve the sensitivity of such searches significantly, in particular offering the possibility of detection of a population of pulsars in the Galactic bulge that would be a background for Galactic Center searches~\cite{Calore:2015bsx,Berteaud:2020zef}. Recent studies of TeV halos around nearby pulsars suggest that TeV gamma-ray observations may also be able to meaningfully constrain pulsar populations~\cite{Song:2019nrx,Macias:2021boz}. In the future, 3rd generation ground-based gravitational wave telescopes have the potential to detect pulsar populations in the Galactic bulge via their contribution to the Galactic stochastic gravitational wave background~\cite{Calore:2018sbp}. More realistic simulation-based studies of population synthesis and dynamical evolution (including stellar binary interactions, pulsar kicks, and mixed formation scenarios) would be helpful to reduce the theory uncertainties on the expected spatial morphology and luminosity function of any population of millisecond pulsars in the inner Galaxy. 

Black holes could serve as DM candidates themselves, or as celestial laboratories for DM searches. These possibilities are discussed in dedicated Snowmass White Papers~\cite{Bird:2022wvk, SnowMassWPxtremenviron}, so we will only mention them briefly here. Gravitational wave searches can probe the population of black hole mergers and thus constrain the primordial black hole abundance, in particular by exploiting information on the spin, eccentricities, and mass spectrum (see Ref.~\cite{CF1WP7} and references therein). Measurements of the stochastic gravitational wave background could also constrain a high-redshift contribution from primordial black holes, or their formation mechanism~\cite{Mandic:2016lcn,Nakama:2016gzw, Bartolo:2016ami}. For primordial black hole searches, astrophysical black holes act as a background, and thus studies of the formation, evolution and signatures of astrophysical black holes may have implications for DM searches. Primordial black hole DM could also lead to implosion of neutron stars~\cite{Fuller:2017uyd}, and thus improved observations of the neutron star population could provide relevant constraints.

\subsubsection*{General theory and analysis needs}

Applying this rich ensemble of multiwavelength and multimessenger observations to elicit information about DM physics will require dedicated efforts with regard to theory, modeling, and computation.

To quote from~\cite{CF1WP7}, {\it ``The ongoing development of open-source tools and the definition of standards that facilitate the sharing of data will be crucial to fully exploit the synergies between different wavelengths and astrophysical messengers''.} There are already some packages and frameworks of this type, such as Gammapy and the Multi-Mission Maximum Likelihood (3ML) framework~\cite{Vianello:2015wwa}, which could serve as a basis for future work in this direction.

The characterization of spatial structure at different scales, including the detection of point sources and separation of point sources from diffuse emission, is a generic problem in a number of indirect searches. A number of analysis techniques have been developed to handle the regime where one cannot characterize individual sources but must instead study the population as a whole. Such strategies include study of the angular power spectrum~\cite{Ando:2005xg, Ando:2006cr,Fermi-LAT:2012pez,Fornasa:2016ohl, Fermi-LAT:2018udj}, probabilistic cataloguing~\cite{Portillo_2017,Feder_2020}, pixel count statistics with non-Poissonian template fitting~\cite{Malyshev:2011zi,Lee:2014mza,Lee:2015fea,Mishra-Sharma:2016gis,IceCube:2019xiu} and the Compound Poisson Generator framework~\cite{Collin:2021ufc}, use of trained neural networks and likelihood-free inference~\cite{Caron:2017udl,List:2020mzd,List:2021aer,Mishra-Sharma:2021oxe}, wavelet decomposition of images to identify structures at different scales~\cite{1992tlw..conf.....D, 2006Stack,Bartels:2015aea, Zhong:2019ycb}, and cross-correlations between related datasets, \eg~\cite{Ando:2013xwa,Ando:2014aoa,Camera:2012cj,Camera:2014rja,Cuoco:2015rfa, Cuoco:2017bpv,DES:2019ucp}. Cross-correlation in particular may benefit from simultaneous progress in expanding several related datasets; for example, the Vera C. Rubin Observatory will increase sky coverage of existing lensing maps by some factor $\sim10$ or more, improving the prospects for correlations between gamma-ray signals and lensing maps~\cite{CF1WP7}. Many of these methods are at a relatively early stage and will benefit from further development.

For ultraheavy DM, a detailed understanding of propagation effects can become important even for photons, as sufficiently high-energy photons can pair produce on the intergalactic radiation field. Furthermore, for DM masses far above the weak scale, standard event generators (such as {\tt Pythia}~\cite{Sjostrand:2006za,Sjostrand:2007gs,Sjostrand:2014zea}) become insufficient to model the production of stable particles from decays of the initial unstable decay/annihilation products. Some theoretical work has already been done on this front~\cite{Bauer:2020jay}, but further work will be required to ensure the availability of accurate and precise predictions for how heavy DM should appear in our telescopes.

\subsubsection{Paths to resolving current puzzling excesses}
\label{sec:indirect-excesses}

{\it This section draws heavily on the conclusions of WP6~\cite{CF1WP6}.}

Many tantalizing excesses have been reported across DM indirect detection experiments. These excesses may or may not hold clues to DM physics, but -- in that they are currently unresolved -- they signal areas where our understanding of background and/or signal uncertainties requires improvement in order to either confirm or exclude a DM interpretation. Here we discuss how the advances discussed above would provide a path toward resolving the origin of these excesses.

\subsubsection*{Galactic Center excess in GeV-scale gamma rays}

A GeV-scale excess at the heart of our Galaxy has been detected in gamma rays by the Fermi Large Area Telescope (Fermi-LAT), with high statistical significance. Leading explanations are either weak-scale annihilating DM, or a new population of gamma-ray emitting pulsars. The key barrier to understanding the nature of the excess is obtaining an accurate model for the Galactic diffuse gamma-ray foreground -- this component makes up the bulk of the gamma rays in the region, but is not well understood. The measurements and theory advances discussed above to improve our understanding of cosmic-ray propagation, Galactic diffuse gamma rays, and multiwavelength/multimessenger pulsar searches provide a path toward resolving the origins of this excess. In the event that the excess does signal new DM physics, improving sensitivity for searches for counterpart signals will also be essential in confirming the detection.

\subsubsection*{Excess in GeV-scale antiprotons}

An excess of GeV cosmic-ray antiprotons in the Alpha Magnetic Spectrometer (AMS-02) observations has been identified. Interpreted as a signal of DM annihilation, this excess requires similar mass,  cross section and annihilation channels to those required to explain the Galactic Center excess. To establish the robustness of this excess, the underlying correlations of the AMS-02 systematic errors are needed. The observation of antideuteron or antihelium nuclei would be an unambiguous signal of new physics. Antiproton and antinuclei DM searches will benefit from future production cross section measurements from inelastic hadronic collisions at low center-of-mass energies, and from further reduction of the astrophysical cosmic-ray propagation modeling uncertainties. 

\subsubsection*{Excess in TeV-scale positrons}

Observations of a rising positron fraction at high energies (tens-hundreds of GeV) have sparked considerable interest over the past fifteen years due to their potential DM explanations. However, high-energy gamma-ray observations over the last few years have produced significant evidence that astrophysical $e^+e^-$ acceleration by a population of high-energy pulsars is the most likely explanation for the excess positron flux observed at Earth. In closing this mystery, these TeV halos have added new questions, as their morphology indicates that the simple picture of isotropic and homogeneous particle diffusion throughout the Milky Way is violated on moderate scales. Understanding this new phenomenon may play an important role in DM indirect detection searches over the next decade, using both positrons as well as cosmic-ray and gamma-ray probes.

\subsubsection*{Excess in 511 keV gamma rays}

The 511 keV line from the decay of non-relativistic positronium coming from the Galactic center has been observed for over 40 years. In particular, the spectrometer on the International Gamma-Ray Astrophysics Laboratory (INTEGRAL/SPI) has measured a nearly spherical morphology for the signal. Various non-conventional DM scenarios have been proposed to explain the excess, which can be complementarily tested in accelerators and astrophysical observations. More detailed understanding of the morphology of the signal is crucial in identifying the origin of the excess.

\subsubsection*{Excess in 3.5 keV X-rays}

The 3.5 keV line is an anomaly detected in the X-ray band that has been interpreted as a possible hint of decaying DM. First observed in the datasets of High Throughput X-ray Spectroscopy Mission X-ray Multi-Mirror Mission (XMM- Newton) and Chandra in 2014, the anomaly exhibited several properties expected of DM, although possible astrophysical explanations such as charge exchange or a potassium emission line were also discussed. Challenges to the simplest DM interpretations have arisen from non-observations, particularly from the halo of the Milky Way. Future instruments in the X-ray band should have sufficiently improved sensitivity to the signal to largely resolve the debate; in particular, upcoming instruments with excellent energy resolution may be able to resolve the line width expected under the DM hypothesis.

\subsection{Experiments, Facilities, and New Technologies}

\begin{figure}
    \centering
    \includegraphics[width=\textwidth]{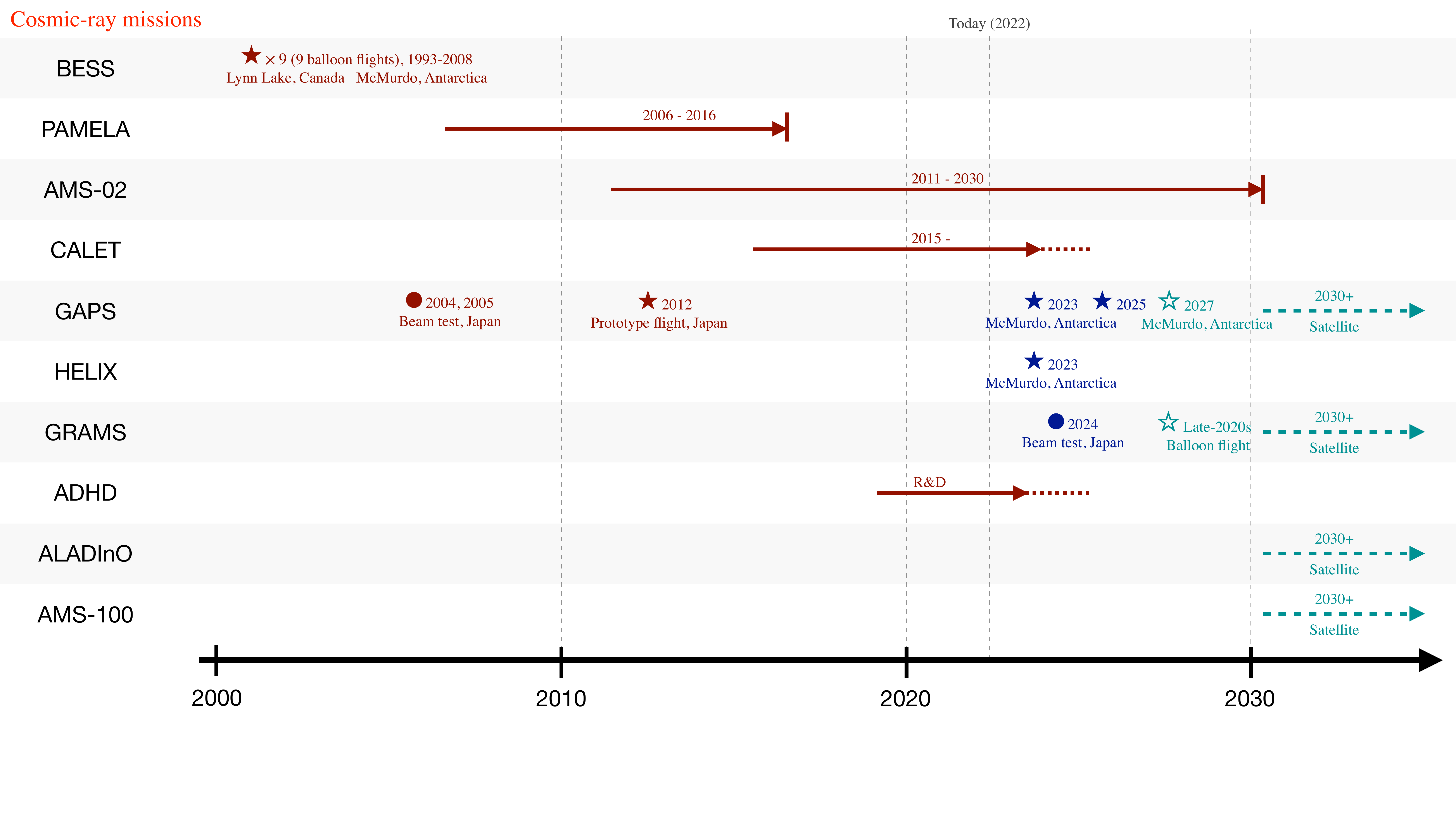}
    \includegraphics[width=\textwidth]{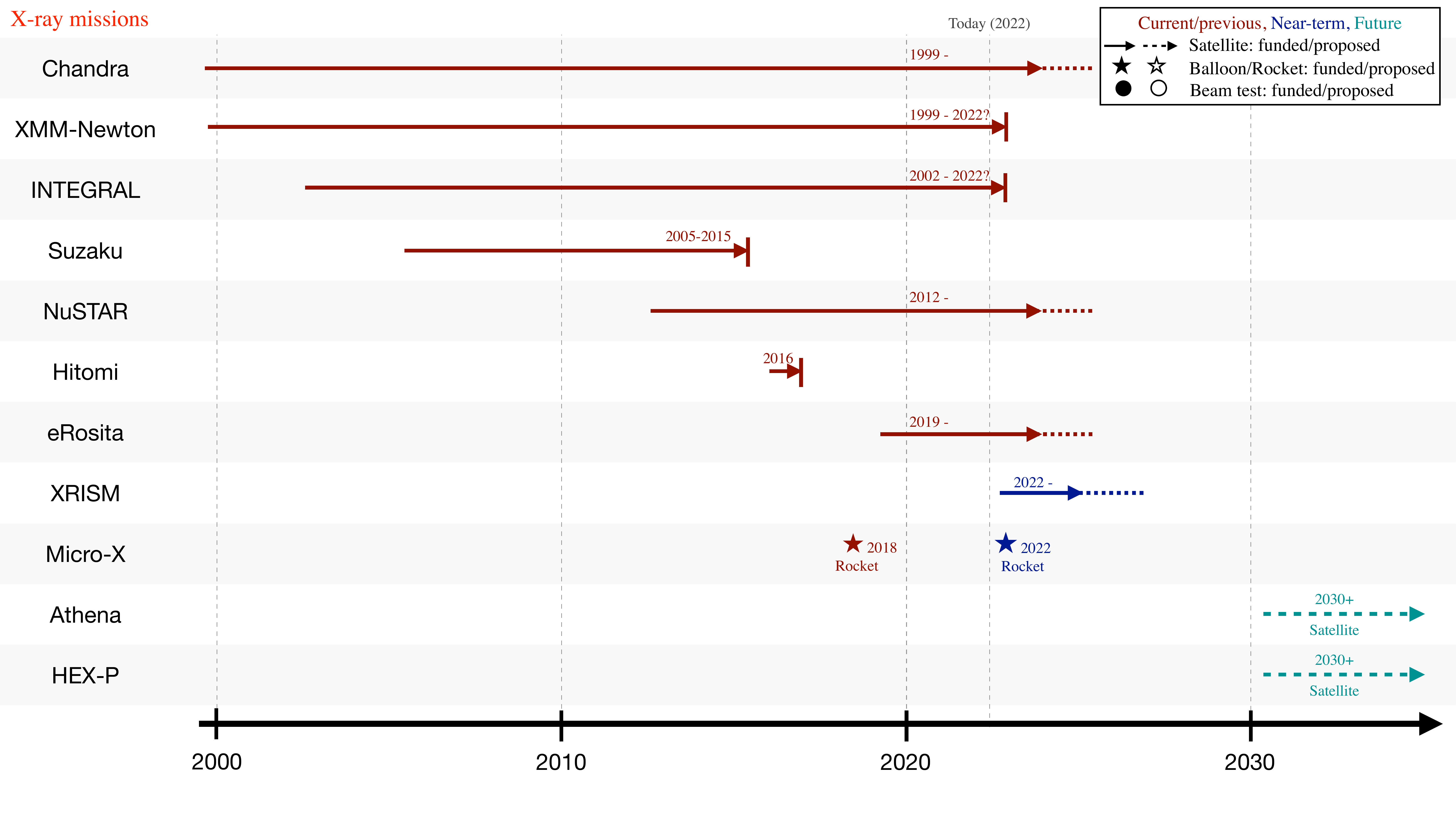}
    \caption{Overview of current, upcoming and proposed missions for cosmic rays (upper panel) and X-rays (lower panel). Current/previous, near-term, and further future missions are marked in red, dark blue and teal respectively. Solid lines/symbols indicate funded experiments while dashed lines or empty symbols indicated proposed experiments. Lines indicate satellite or ground-based missions, stars indicate individual balloon flights, and circles indicate beam tests.  Reproduced from Ref.~\cite{CF1WP5}.}
    \label{fig:WP5missions_XRCR}
\end{figure}

\begin{figure}
    \centering
    \includegraphics[width=\textwidth]{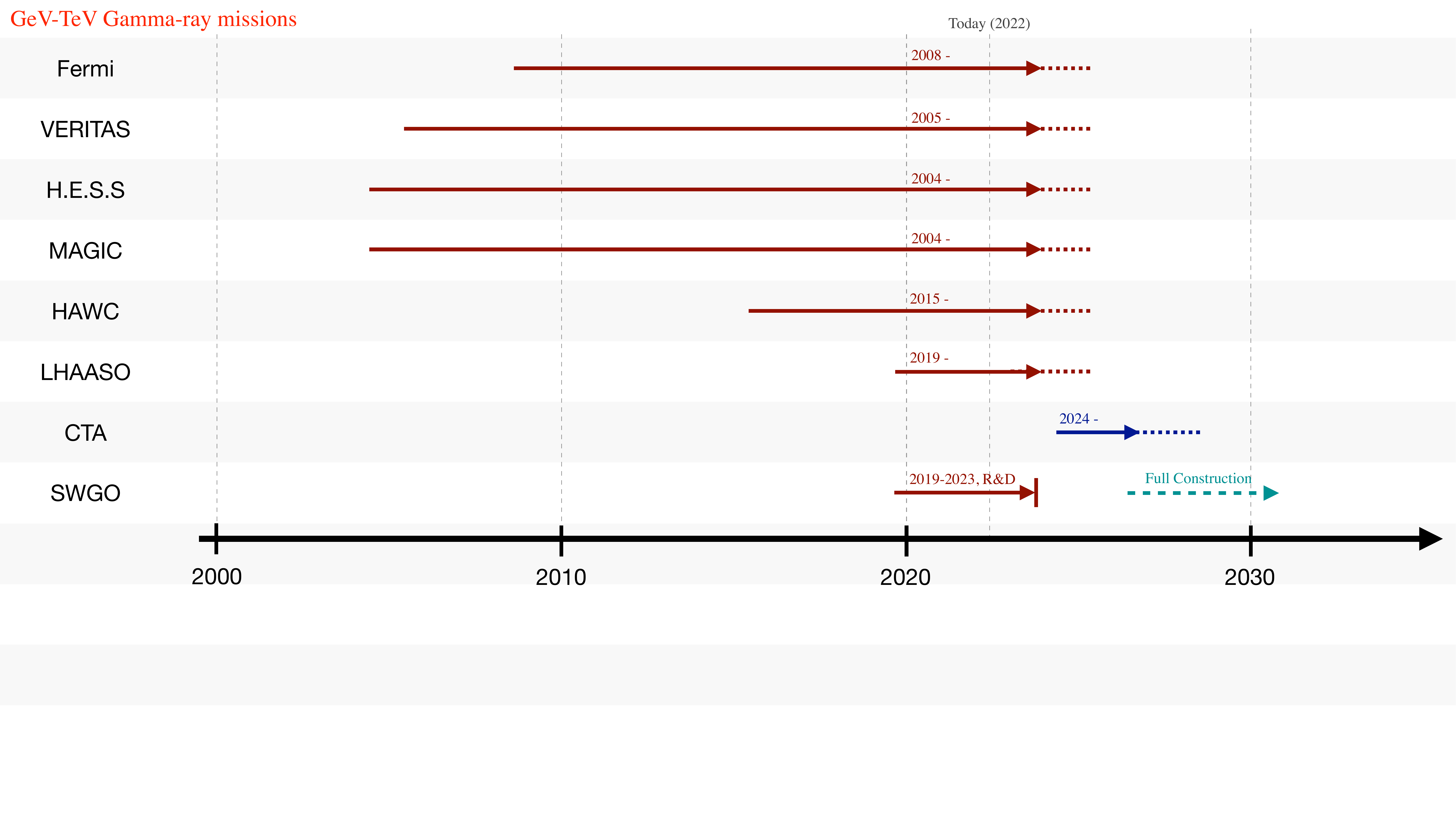}
    \includegraphics[width=\textwidth]{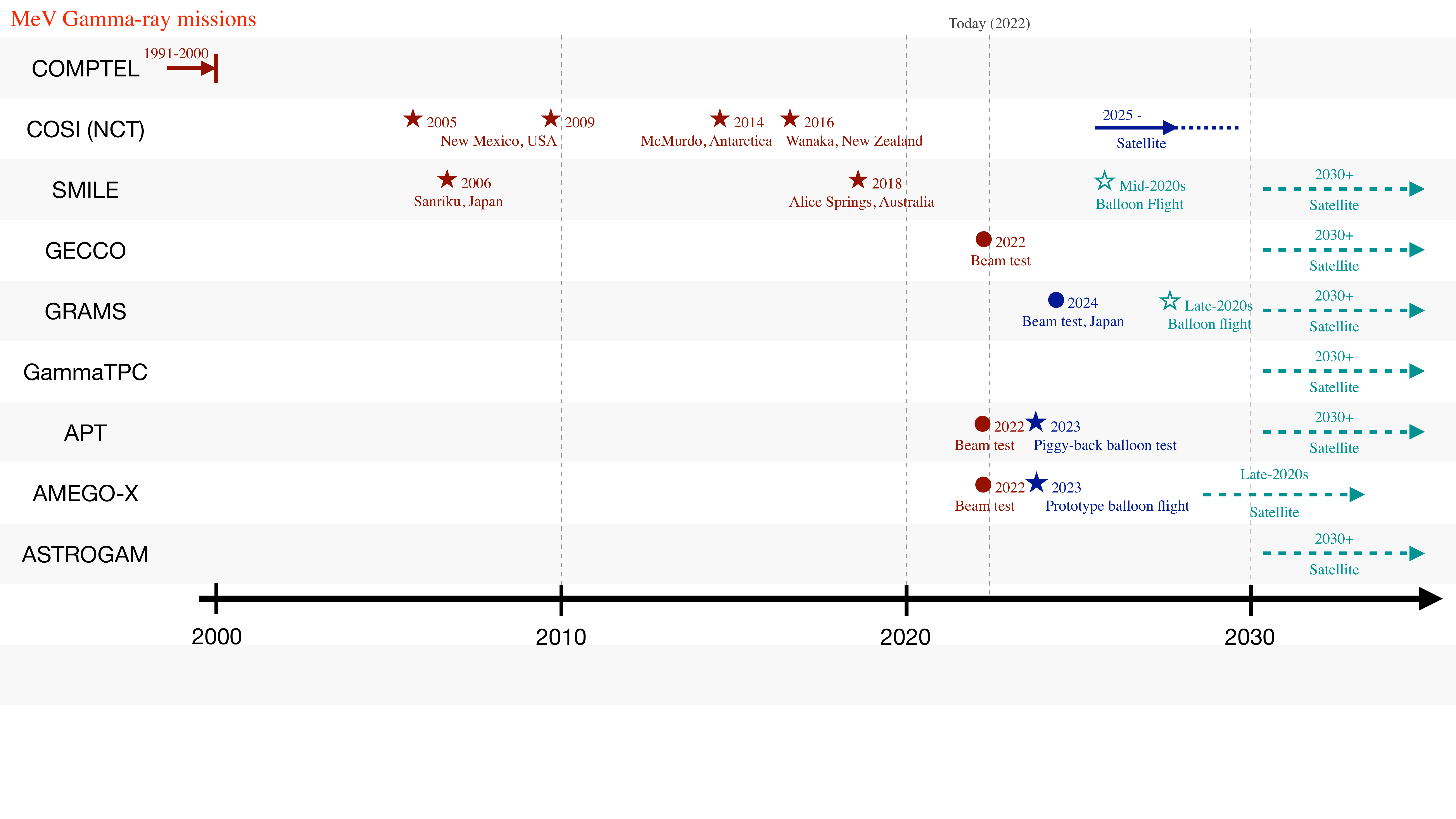}
    \caption{Overview of current, upcoming and proposed missions for gamma-rays in the GeV-TeV+ band  (upper panel) and MeV band (lower panel). Current/previous, near-term, and further future missions are marked in red, dark blue and teal respectively. Solid lines/symbols indicate funded experiments while dashed lines or empty symbols indicated proposed experiments. Lines indicate satellite or ground-based missions, stars indicate individual balloon flights, and circles indicate beam tests. Reproduced from Ref.~\cite{CF1WP5}}
    \label{fig:WP5missions_GR}
\end{figure}

Figures~\ref{fig:WP5missions_XRCR} and \ref{fig:WP5missions_GR} outline current and proposed missions probing DM indirect signals with cosmic rays and gamma rays, while Table~\ref{table-indirect} provides a summary of photon and neutrino experiments sensitive to ultra-heavy DM. Details of the proposed cosmic-ray and gamma-ray probes and their relevance for indirect detection can be found in \cite{CF1WP5}; we provide a brief outline of some key goals of these instruments and how they relate to the needed measurements discussed above. The full multimessenger observational program, and its relevance for probes of fundamental physics including but not limited to DM, is discussed in the CF7 white papers \cite{Engel:2022bgx,Engel:2022yig}.

    \begin{table}[t]
\begin{tabular}{|l|l|l|l|l|l|}
\hline
Experiment & Final state & Threshold/sensitivity & Field of view & Location  \\
\hline
\hline 
\multicolumn{5}{|c|}{Current experiments} \\
\hline
Fermi & Photons & $10~{\rm MeV} -10^{3}~{\rm GeV}$ & Wide & Space  \\
\hline
HESS & Photons & 30 GeV - 100 TeV & Targeted & Namibia  \\
\hline
VERITAS & Photons & 85 GeV - $>$ 30 TeV & Targeted & USA  \\
\hline
MAGIC & Photons & 30 GeV - 100 TeV & Targeted & Spain  \\
\hline
HAWC & Photons & 300 GeV - $>$100 TeV  & Wide & Mexico  \\
\hline
LHAASO (partial) &  Photons & 10 TeV - 10 PeV & Wide &  China  \\
 \hline
KASCADE &  Photons & 100 TeV - 10 PeV & Wide &  Germany  \\
 \hline
KASCADE-Grande &  Photons & 10 - 100 PeV & Wide &  Italy  \\
 \hline
Pierre Auger Observatory &  Photons & 1 EeV - 1 ZeV & Wide &  Argentina  \\
 \hline
Telescope Array &  Photons & 1 - 100 EeV  & Wide &  USA  \\
 \hline
IceCube & Neutrinos & 100 TeV - 100 EeV & Wide & Antarctica   \\
\hline
ANITA & Neutrinos & EeV - ZeV & Wide &  Antarctica \\
\hline
Pierre Auger Observatory &  Photons \& Neutrinos & 1 EeV - 1 ZeV & Wide &  Argentina  \\
\hline
\hline
\multicolumn{5}{|c|}{Future experiments} \\
\hline
CTA & Photons & 20 GeV - 300 TeV & Targeted & Chile \& Spain  \\
 \hline 
SWGO & Photons & 100 GeV - 1 PeV & Wide & South America  \\
 \hline 
IceCube-Gen2 & Neutrinos & 10 TeV - 100 EeV & Wide & Antarctica  \\
\hline
LHAASO (full) &  Photons & 100 GeV - 10 PeV & Wide &  China  \\
 \hline 
KM3NeT & Neutrinos & 100 GeV - 10 PeV & Wide  & Mediterranean Sea  \\
 \hline 
 AugerPrime &  Photons \& Neutrinos & 1 EeV - 1 ZeV & Wide &  Argentina  \\
 \hline
POEMMA & Neutrinos & 20 PeV - 100 EeV & Wide &  Space  \\
 \hline
\end{tabular}
\caption{A non-exhaustive list of current and future indirect detection experiments sensitive to ultraheavy DM. See Refs.~\cite{Ishiwata:2019aet,KASCADEGrande:2017vwf,PierreAuger:2015fol,PierreAuger:2016kuz,TelescopeArray:2018rbt,IceCube:2016uab,Kopper:2017zzm,ANITA:2019wyx,Veberic:2017hwu}. Based on table from Ref.~\cite{CF1WP8}.}
\label{table-indirect}
\end{table}

\subsubsection{Improving sensitivity to dark matter signals in high-energy gamma rays: the thermal relic benchmark and beyond}

At high energies, the space-based Fermi-LAT and ground-based atmospheric and water Cherenkov telescopes have set stringent limits on DM annihilation and decay, including probing the ($s$-wave) thermal relic cross section across a range of annihilation channels for DM masses as high as tens to hundreds of GeV. Current imaging atmospheric Cherenkov telescopes (IACTs) include VERITAS, HESS, and MAGIC; these instruments operate by measuring Cherenkov light produced in the atmosphere due to showers initiated by astrophysical $\upgamma$-rays and charged cosmic rays. Water Cherenkov detectors, including HAWC and LHAASO, use water tanks to detect charged particles generated by high-energy $\upgamma$-ray or cosmic-ray interactions in the atmosphere. These approaches are complementary: the water Cherenkov detectors have much wider fields of view, allowing for all-sky searches, transient searches, and stringent constraints on extended and nearby sources, while IACTs achieve higher instantaneous sensitivity by virtue of lower energy thresholds, improved angular resolution, and better identification of the primary particle type. At energies below $\sim$ 100 GeV, these approaches lose effective area and space-based probes are required; the current flagship instrument for indirect DM searches in the $1-100$ GeV range is the Fermi-LAT, which has full-sky coverage and operates via pair conversion of incident gamma rays in tracker layers.

At 100 GeV and higher energies, the key upcoming and proposed experiments are CTA and the Southern Wide-Field Gamma-Ray Observatory (SWGO), which will use IACT and water Cherenkov technology respectively. 

With installations in both the Northern and Southern Hemispheres (Spain and Chile), CTA aims to improve sensitivity by a factor between five and twenty (depending on the energy) compared to current generation IACTs, for gamma-ray energies between 20 GeV and 300 TeV. CTA expects to probe the thermal annihilation cross section for DM masses in the range $\sim$ 0.2--20 TeV, complementing existing constraints from Fermi-LAT at lower energies; extensive studies of CTA's potential for DM searches have been performed by the CTA Consortium~\cite{2019scta.book.....C, CTA:2020qlo}. In addition to testing the thermal relic benchmark (and setting broader constraints on annihilation and decay) via gamma-ray observations, CTA can constrain the cosmic ray electron-positron spectrum up to hundreds of TeV~\cite{2019scta.book.....C,ctaelectrons}, and can conduct complementary observations of TeV halos around nearby pulsar wind nebulae. As discussed above, a better understanding of pulsars at all energy scales would be valuable for resolving current excesses and for background modeling in general. 

The reach of these searches with CTA may benefit from further developments in telescope technology; for example, an international consortium of CTA members, led by the U.S., has developed and prototyped a novel medium-sized telescope design for CTA, called the Schwarzschild-Couder Telescope (SCT)
~\cite{2007APh....28...10V,2008ICRC....3.1445V, Adams:2020qxs}, to both augment the telescope count and improve the angular resolution of the instrument, allowing for improved background rejection. Currently funding has been identified and committed to building an ``Alpha Configuration'' of CTA during 2022--2027; while exceeding   the capabilities of any of the existing IACT arrays, this configuration would have (for funding reasons) fewer telescopes than initially envisioned for CTA. However, adding $\sim 10$ SCTs to this configuration in the south would enhance CTA's capabilities for DM searches in the Galactic Center region to at least the level originally anticipated (and employed in the studies mentioned above).

SWGO is planned to be located in the Southern Hemisphere, and to have sensitivity exceeding that of HAWC by a factor of $\sim 10$. The Southern Hemisphere location is advantageous for observations of the Galactic Center and the central dense region of the Milky Way's DM halo; both HAWC and LHAASO are Northern Hemisphere installations. The ability to observe the Galactic Center will enable SWGO to search for DM signals 2-3 orders of magnitude fainter than HAWC. With its large field of view, SWGO can conduct a wide range of DM searches. In particular, the Rubin Observatory~\cite{Mandelbaum:2018ouv} will survey the Southern Hemisphere sky with unprecedented sensitivity and is expected to find hundreds of new dwarf spheroidals ~\cite{Hargis:2014kaa}; all-sky data from SWGO will allow immediate follow-up of newly-discovered dwarf galaxies. SWGO observations of the Galactic halo and the Galactic center have been forecast to probe the thermal relic cross section for masses up to tens of TeV for most possible final states~\cite{Viana:2019ucn}.

SWGO will also expand sensitivity to ultra-heavy decaying DM; at present, for the example annihilation channel of DM annihilating to $b$ quarks (which should be broadly similar to other channels involving hadronization) Fermi-LAT and HAWC can probe lifetimes as long as $\mathcal{O}(10^{27-28})$ s for masses up to $\mathcal{O}(10^6-10^7)$ GeV; SWGO will probe lifetimes in the $10^{28-30}$ s range for DM masses as high as $10^8$ GeV~\cite{CF1WP8}. As indicated in Table~\ref{table-indirect}, LHAASO and planned neutrino experiments (IceCube-Gen2, KM3NeT, POEMMA) will also provide sensitivity to ultraheavy DM in the coming years.

At somewhat lower energies, the Advanced Particle-astrophysics Telescope (APT)~\cite{buckleyastro2020} is a concept for a future space-based successor instrument to the Fermi-LAT, with a nearly all-sky field-of-view and the sensitivity to probe thermal relic dark matter up to the TeV scale. The ADAPT (Antarctic Demonstrator for APT) suborbital mission was recently funded by NASA and is planned for a 2025 Antarctic flight. APT would aim to improve on the sensitivity of Fermi-LAT by a factor of $\sim 10$ in the 100 MeV--TeV energy range  while also performing dramatically better in the 10-100 MeV band where the Fermi-LAT loses sensitivity.

\begin{figure}[!htbp]
\begin{center}
\includegraphics[width=0.68\columnwidth]{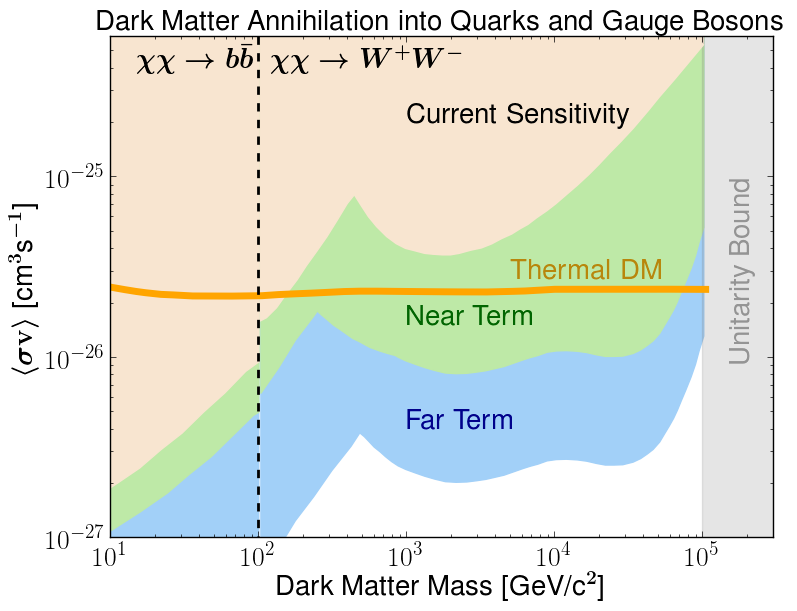}
\caption{Current and expected sensitivity to $s$-wave DM annihilation to quarks and gauge bosons, achieved by space-based and ground-based high-energy gamma-ray experiments, for both current (beige) and future (green, blue)
searches. Cross section values are extracted from Refs.~\cite{Fermi-LAT:2016uux,Fermi-LAT:2019lyf,HESS:2022ygk,Viana:2019ucn,CTA:2020qlo}, as well as far-future sensitivities as described above.
}
\label{fig:thermalxsec}
\end{center}
\end{figure}

Fig.~\ref{fig:thermalxsec} shows the sensitivity of current, near-future, and far-future searches for DM annihilation to representative channels $b\bar{b}$ ($W^+W^-$) for DM masses below (above) 100 GeV, assuming an Einasto profile for the Galactic halo \cite{Einasto}. The near-future band takes into account the enhanced sensitivity of the Fermi-LAT in the event of the discovery of additional dwarfs with LSST~\cite{Fermi-LAT:2016uux}, as well as forecasts from CTA~\cite{CTA:2020qlo} and SWGO~\cite{Viana:2019ucn}, and has the capacity to probe thermal relic DM up to the $\mathcal{O}(100)$ TeV DM mass range where unitarity begins to exclude thermal freezeout as the mechanism fixing the DM relic density in a standard cosmology (the details of the exact unitarity-based mass bound are dependent on \eg the role played by bound states, whether the DM is self-conjugate, and whether it is composite \cite{Smirnov:2019ngs}). The far-future band assumes successful operation of APT improving on the Fermi-LAT sensitivity by a factor of 10, as well as a SWGO successor experiment with km$^2$ area, and would offer sensitivity to the thermal target even in the event of a flattened density profile for the Galactic halo.

\subsubsection{Improving sensitivity to dark matter signals in the MeV band: closing the MeV gap}

The last major experiment in the MeV-100 MeV gamma-ray band was NASA's Imaging Compton Telescope (COMPTEL), which operated from 1991-2000. COMPTEL relied on measuring energy deposition from gamma rays Compton scattering in the detector, and measured a range of Galactic and extragalactic sources of gamma-ray emission. X-ray and soft gamma-ray telescopes have since probed the sub-MeV energy range with greater sensitivity, and the Fermi-LAT performs well at energies of 100 MeV and higher, leaving a ``sensitivity gap'' in this energy range. As discussed in Sec.~\ref{sec:indirect-complementary}, closing this gap is important for enabling data-driven studies of backgrounds at both lower and higher energies, as well as providing greater sensitivity to light DM in the MeV-GeV mass range.

In the near term, the Compton Spectrometer and Imager (COSI) has a planned launch date in 2025, and will survey the gamma-ray sky at energies of 0.2-5 MeV~\cite{Tomsick:2021wed} with a large field of view and excellent energy resolution. COSI will significantly improve current constraints on decaying or annihilating light DM; for annihilation directly to monochromatic photons in the Galactic Center, its (five sigma discovery) reach in annihilation cross section can be estimated as $3.5\times10^{-35}\,$cm$^3$/s, $1.4\times10^{-34}\,$cm$^3$/s, and $1.3\times10^{-33}\,$cm$^3$/s, for DM masses 0.3\,MeV, 1\,MeV, and 3\,MeV, respectively~\cite{CF1WP5}.

Proposed future experiments in this energy range include the Sub-MeV $\upgamma$-ray Imaging Loaded-on balloon Experiment (SMILE)~\cite{takada2021first,tanimori2020mev,hamaguchi2019space,takada2020smile}, Galactic Explorer with a Coded Aperture Mask Compton Telescope (GECCO)~\cite{moiseevsnowmass2021}, Gamma-ray and AntiMatter Survey (GRAMS)~\cite{Aramaki:2019bpi}, Gamma Time Projection Chamber (GammaTPC), APT (as discussed above), All-sky Medium Energy Gamma-ray Observatory eXplorer (AMEGO-X)~\cite{Fleischhack:2021mhc}, and enhanced ASTROGAM (e-ASTROGAM)~\cite{e-ASTROGAM:2017pxr}. These proposed instruments explore a range of new technologies for gamma-ray detection; details may be found in Ref.~\cite{CF1WP5}, but as one example, SMILE, GRAMS and GammaTPC rely on gaseous or liquid time projection chambers (TPCs), with synergy with significant developments in recent years on liquid noble TPCs for DM and neutrino physics. GECCO is based on a novel CdZnTe imaging calorimeter with a deployable coded aperture mask. AMEGO-X would employ a silicon pixel tracker, CsI calorimeter, and plastic anticoincidence detector to track both Compton interactions and $e^\pm$ pair production; e-ASTROGAM would similarly rely on a silicon tracker combined with a calorimeter, resulting in a sensitivity one to two orders of magnitude better than that of COMPTEL.

All these mission concepts would improve on existing experiments via a combination of energy resolution, angular resolution, and effective area. For example, the GECCO experiment expects to achieve sub-percent energy resolution and arcminute angular resolution under some circumstances, compared to a $1-2^\circ$ angular resolution and $9\%$ energy resolution for COMPTEL~\cite{1993ApJS...86..657S}. Most of these instruments are focused on energies around 0.1-10 MeV, but the APT would retain adequate energy reconstruction up to TeV energies as discussed above, and AMEGO-X will probe energies from 25 keV--1 GeV.

As a result of these improvements, these missions would significantly improve on current sensitivity from indirect searches to annihilation and decay of DM in this mass range, especially for photon-rich final states (including production of neutral pions, as well as annihilation/decay to $\gamma+X$) \cite{CF1WP5, Coogan:2021sjs}. There are existing stringent limits from the cosmic microwave background (CMB) for $s$-wave annihilation \cite{Planck:2018vyg}, but present-day indirect-detection experiments offer unique sensitivity to $p$-wave annihilation that is strongly suppressed due to low DM velocities during the CMB epoch, as well as decaying DM (e.g.~\cite{CF1WP5,Cirelli:2020bpc}). Models of light thermal DM with $p$-wave annihilation are frequently favored benchmarks for terrestrial experiments, and future indirect-detection experiments in the MeV band could potentially reach thermal relics annihilating dominantly through $p$-wave processes \cite{Coogan:2021sjs}.

\subsubsection{Improving sensitivity to dark matter signals in X-rays}

Observations performed with the current generation of X-ray instruments provide the leading direct experimental constraints on sterile neutrino decay over the complete keV-mass range, where sterile neutrinos could constitute DM~\cite{PhysRevLett.72.17},  and leading limits on axion-like particles for masses $<10^{-11}$\,eV. These experiments include Chandra~\cite{2000SPIE.4012....2W},  XMM-Newton ~\cite{2001A&A...365L...1J},  INTEGRAL~\cite{KUULKERS2021101629}, Suzaku~\cite{2007PASJ...59S...1M}, and the Nuclear Spectroscopic Telescope Array (NuSTAR)~\cite{Harrison:2013, Madsen:2015}. The sterile neutrino bounds are highly complementary with limits on warm dark matter from cosmic probes~\cite{Bechtol:2022koa}, as discussed in Ref.~\cite{CF1WP5}.

Upcoming experiments offer the promise of improved sensitivity and energy resolution; the latter factor is especially important for identifying monochromatic line signals and distinguishing them from possible backgrounds, and should allow an observational resolution of the claimed 3.5 keV line excess. In particular, energy resolution at the $10^{-3}$ level should permit a measurement of the Doppler broadening associated with the velocity of DM particles in our Galaxy. 

In the near term, the Extended ROentgen Survey with an Imaging Telescope Array (eROSITA)~\cite{Sunyaev2021}, which launched in 2019, will provide enhanced sensitivity to weak X-ray lines and spectral distortions due to ultralight axions interconverting with photons. In the next few years, X-ray Imaging and Spectroscopy Mission (XRISM)~\cite{Tashiro2020,XRISM2020} and Micro-X~\cite{2020JLTP..199.1072A} will target fine energy resolutions of 7 eV and 3 eV respectively, in the 0.5-10 keV energy range. 

In the longer term, the next-generation X-ray observatory Athena~\cite{Nandra2013}, selected 
by the European Space Agency, is scheduled to launch in the early 2030s and will carry two 
X-ray instruments, one optimized for a large field of view and the other for excellent energy
resolution (2.5 eV). The latter instrument will have an effective area exceeding that of 
XRISM by at least an order of magnitude at 1 keV. HEX-P~\cite{Madsen2019HEX} is a proposed 
next-generation X-ray observatory that would also target a launch date in the early 2030s, 
and would complement Athena's low-energy reach at high energies (2--200 keV). With 40\,times
the sensitivity of any previous mission in the 10--80\,keV band, and $>$100 times the 
sensitivity of any previous mission in the 80--200\,keV band, HEX-P would have unprecedented
sensitivity to secondary photon signals from $\sim$1--100s MeV particle dark matter decaying
or annihilating into $e^+e^-$, $\mu^+\mu^-$, or pions.

\subsubsection{Improving sensitivity to dark matter signals in cosmic rays}

There is a long history of DM searches in charged cosmic rays, by experiments including the Balloon-borne Experiment with Superconducting Spectrometer (BESS), Payload for Antimatter Matter Exploration and Light-nuclei Astrophysics (PAMELA), AMS-02, Calorimetric Electron Telescope (CALET), and Dark Matter Particle Explorer (DAMPE). Measurements of matter and antimatter cosmic ray spectra by these instruments have  constrained cosmic-ray propagation in our Galaxy, set strong limits on DM annihilation and decay, and in some cases revealed surprising features (some of which are discussed in Sec.~\ref{sec:indirect-excesses}).

In the near-term future, the High Energy Light Isotope Experiment (HELIX)~\cite{f2f8e111be344c99b12978ac18b696ce} instrument is preparing for its first flight as a high-altitude balloon payload. The primary goal of HELIX is to measure the ratio of Beryllium-10 to Beryllium-9 ratio in the energy range of about 1--10\,GeV/$n$, which will provide a measurement of the diffusion time of cosmic rays in our Galaxy and hence provide important constraints on cosmic-ray propagation.

Longer term, there are proposals for two next-generation magnetic spectrometers with significantly increased acceptance when compared to AMS-02: the Antimatter Large Acceptance Detector In Orbit (ALADInO)~\cite{Battiston:2021org,instruments6020019} and A Magnetic Spectrometer (AMS-100)~\cite{2019NIMPA.94462561S}. These detectors could significantly improve on particle-antiparticle separation and on existing measurements of cosmic rays and gamma rays.

\begin{figure}
    \centering
\includegraphics[width=1.\columnwidth]{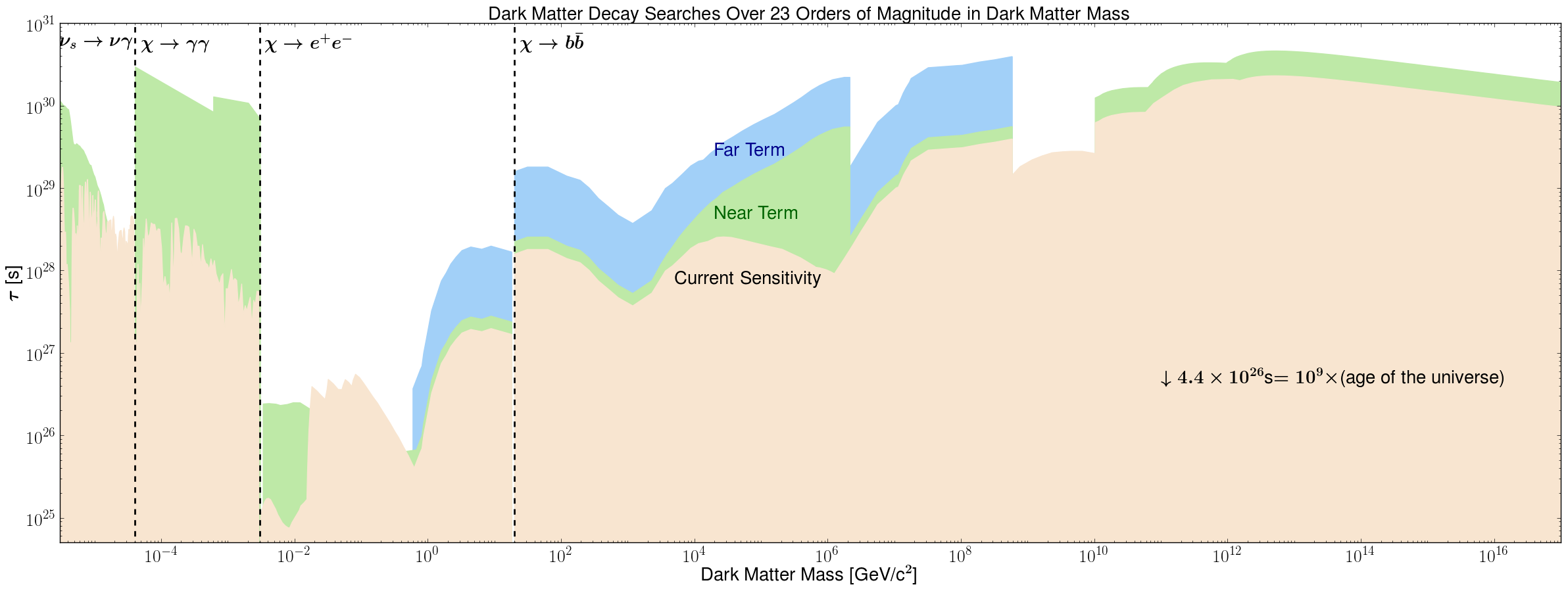}
    \caption{Current and expected sensitivity to DM decay, achieved by cosmic-ray and gamma-ray experiments, for both current (beige) and future (green, blue)
searches. Over 23 orders of magnitude in DM mass can be searched, from sterile neutrino decays at keV energies to DM decaying to gamma rays (40 keV-MeV), electrons and positrons (MeV-10 GeV), and bottom quarks (\textgreater10 GeV). Lifetime values are based on results in Refs.~\cite{Neronov:2015kca,Boyarsky:2007ge,Roach:2022lgo,Foster:2021ngm,Horiuchi:2013noa,Dekker2021,Barinov2021,Bartels:2017dpb,Boyarsky:2007ge,Tomsick:2021wed,Laha:2020ivk,Boudaud:2016mos,Fleischhack:2021mhc,Cohen:2016uyg,Liu:2020wqz,Boudaud:2016mos,buckleyastro2020,Fleischhack:2021mhc,HAWC:2017udy,Cohen:2016uyg,Viana:2019ucn,IceCube:2018tkk,Anchordoqui:2021crl,Chianese:2021jke,buckleyastro2020}. Note that the sharp edges of the curves at most masses are due limited mass ranges of published dark matter results, rather than being from instrumental thresholds.}
    \label{fig:dmdecaywide}
\end{figure}
Fig.~\ref{fig:dmdecaywide} shows current and forecast limits on the DM decay lifetime from observations of charged cosmic rays and gamma rays. These measurements allow us to probe DM lifetimes $\sim 9$ orders of magnitude longer than the age of the universe over an enormous range of DM masses, from the keV scale to the Planck scale, demonstrating both the wide reach and unique capabilities of indirect detection experiments. The enhanced sensitivity of future experiments to decays will in general also be reflected in enhanced sensitivity to annihilation.

\subsubsection{Low-energy antinuclei as a background-free discovery channel}

As yet, there has been no confirmed detection of low-energy antideuterons or antihelium (although there are some tentative possible antihelium events observed by AMS-02), and the expected background for these channels is essentially zero. A detection of such low-energy antinuclei would be transformative.

The General Antiparticle Spectrometer (GAPS) experiment~\cite{Aramaki2015} is the first dedicated experiment to search for low-energy ($<0.25$ GeV/$n$) cosmic-ray antinuclei; it is currently preparing for its first Antarctic long-duration balloon flight. The initial program consists of three flights and the GAPS collaboration is planning a future upgraded payload to increase sensitivity by a factor of 5. GAPS will provide the first precision antiproton spectrum measurement at energies below $0.25$\,GeV/$n$, and provide sensitivity to antideuterons that is about two orders of magnitude better than the current BESS limits, while also having competitive sensitivity to antihelium.

The GRAMS experiment, in addition to targeting MeV-band gamma rays as discussed above, would provide a next-generation antideuteron search using liquid argon TPC technology. GRAMS would have sensitivity to DM models that could explain the Fermi Galactic Center excess and the AMS-02 antiproton excess~\cite{Aramaki:2019bpi}, as discussed in Sec.~\ref{sec:indirect-excesses}.

A significant motivation for ALADInO and AMS-100 is to measure the antideuteron spectrum at energies below $\sim 10$ GeV, and to probe antihelium in light of the possible antihelium events observed by AMS-02. New detection techniques may also enable new opportunities for future antideuteron searches. The proposed AntiDeuteron Helium Detector (ADHD) could provide a novel particle identification approach exploiting the delayed annihilation of antinuclei in helium, which might be used standalone or as a subdetector with traditional techniques.

\begin{figure}
    \centering
    \includegraphics[width=0.5\textwidth]{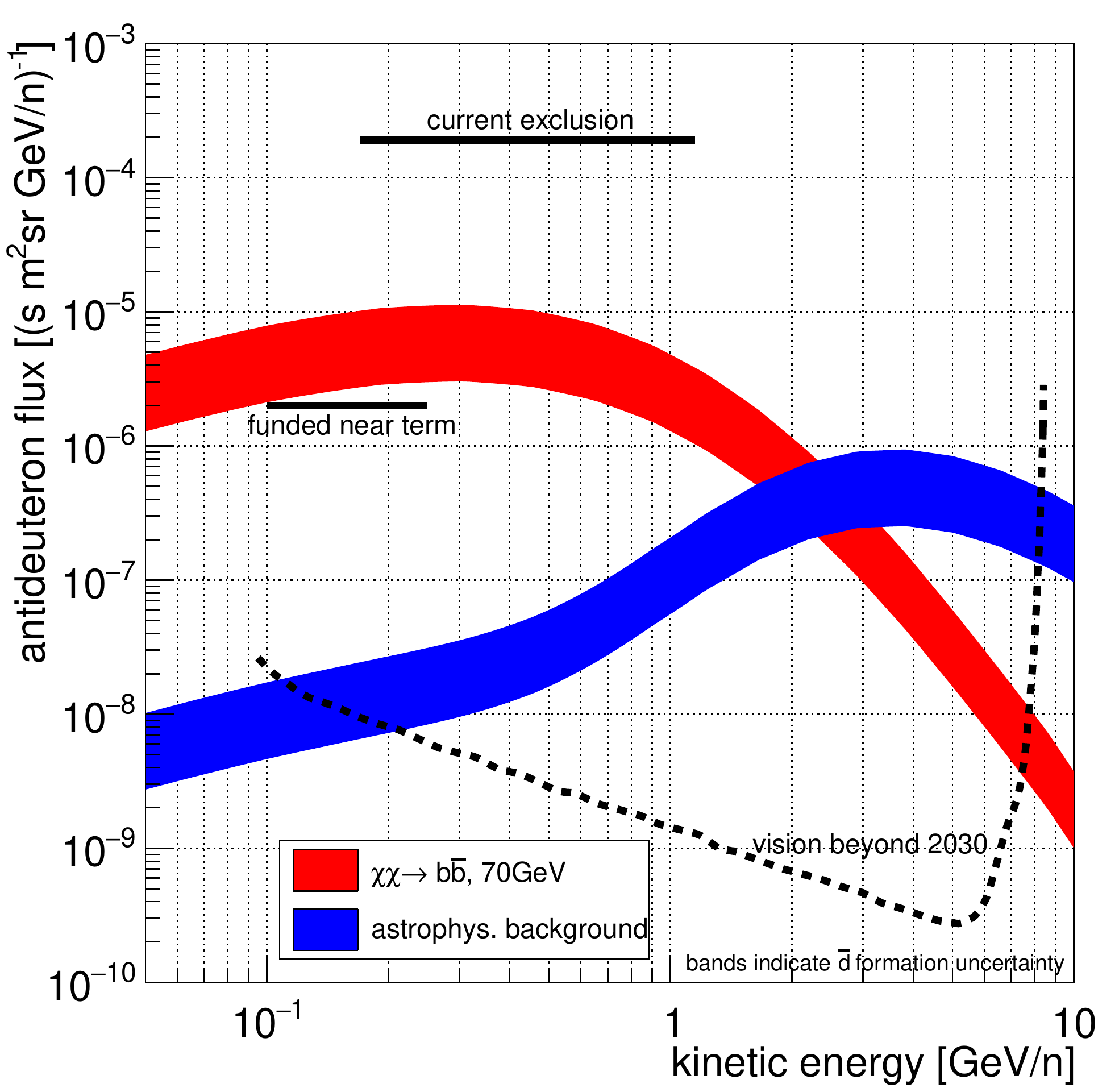}
    \caption{Antideuteron flux from a generic dark matter particle with a mass of $70$\,GeV annihilating into $b\bar b$ compared to the astrophysical background flux prediction, which is orders of magnitude smaller. See Ref.~\cite{CF1WP5} for more details.}
    \label{fig:antinuclei}
\end{figure}
Fig.~\ref{fig:antinuclei} shows the flux sensitivity of current and upcoming experiments to antideuterons. Searches for anti-nuclei, including autideuterons, have been identified as a vital new signature of DM annihilation or decay, essentially free of astrophysical background~\cite{Coogan:2017pwt}. A first-time detection of low-energy cosmic antideuterons would be an unambiguous signal of new physics, complementing collider, direct, and indirect DM searches~\cite{Aramaki:2015pii,vonDoetinchem:2020vbj}. 
\section{Direct Detection}
\subsection{Introduction}
Direct probes of DM are those that search for signatures of local galactic DM particle interactions in terrestrial targets.  These signatures are relatively model independent, relying on low momentum transfers and largely agnostic to the details of the interaction. Traditionally, the signature of choice in this type of detector has been nuclear recoils  -- an elastic collision between the DM particle and a target nucleus.  However, in recent years direct detection searches have expanded to include signatures of inelastic scattering and interactions with electrons.
The typical DM signal is a recoil rate that exceeds the detector-internal background. An annual rate and energy modulation or a daily directional modulation can provide further confirmation.
Because of their agnosticism, many direct detection techniques are capable of searching simultaneously for multiple potential signatures of DM scattering (\eg~DM-nucleon and DM-electron scattering, or thermal DM vs. asymmetric DM) and over a wide range of masses.

The technique of direct detection can probe well motivated DM particle candidates in the meV - 
100 TeV mass scale range.  Discovering a new particle using the technique of direct detection in an underground experiment would constitute evidence that such a particle constitutes all or at least part of the DM.   In addition, in many theoretical models direct detection provides the \textit{only} near-term opportunity for its detection.  Examples of these scenarios include models in which DM particle couples feebly to ordinary matter through a light or massless mediator, and models where interactions between sub-keV bosonic DM and the SM are already strongly constrained to avoid an anomalously large stellar cooling rate \cite{CF1WP2}.

Importantly, direct detection experiments are adaptable, and can respond readily to signals or hints or the lack thereof to mitigate systematic backgrounds or redirect to new parameter space. Furthermore, direct detection experiments can be, and often are, adjusted, upgraded, fixed, or moved to confirm or disprove any observation of DM that may arise.

\subsection{Science targets for direct detection}

We will continue to use the categorization of DM candidates as defined in Sec.~\ref{sec:intro}:  ``light" (sub-GeV), ``heavy'' or ``electroweak scale'' (GeV--100 TeV), and ``ultra-heavy'' ($\gg 100$ TeV).  
The most well-studied DM candidates are in the ``heavy" category; this class of models, which includes WIMPs, gives rise to highly-predictive scattering cross-sections and thus lead to clear benchmark targets for direct detection experiments. 
Examples of these models span a wide range from SUSY models such as the Next-to-Minimal Supersymmetric Models (NMSSM)~\cite{Lopez-Fogliani:2021qpq,Wang_2020} and the Phenomenological Minimal Supersymmetric Models (pMSSM)~\cite{VanBeekveld:2021tgn}, to extensions of the SM which add \eg multiple scalar doublets~\cite{Cabrera:2019gaq,Mukherjee:2022kff}, electroweak Majorana triplets~\cite{Chen:2018uqz}, right-handed neutrinos (RHN)~\cite{Boyarsky:2018tvu}, or pseudo-Goldstone bosons~\cite{Alanne:2020jwx}. Despite the tremendous progress in direct detection experiments, there are still motivated regions of parameter space left unexplored for many of these models~\cite{CF1WP1}. The spin-structure of the DM-nucleon interactions lead to a splitting of the scattering cross-section into Spin-Independent (SI) and Spin-Dependent (SD) scattering. Even for the most predictive models, the relevant DM parameter space is determined by the cosmology during the epoch in which the DM relic abundance was produced. Lack of exact knowledge of the cosmological history of DM compels us to cast a wide net in our DM searches, motivating a diverse experimental program spanning multiple scales and techniques.
 
There is an ensemble of theoretically-motivated ``ultra-heavy" DM (UHDM) candidates, which lead to striking signatures in direct detection experiments due to their mass scales. Non-thermal UHDM candidates, known as WIMPzillas, are a generic phenomenon of inflationary cosmology~\cite{Chung:1998zb,Kolb:1998ki}, whereas $Q$-balls (\eg~\cite{hepph0303065}) provide a candidate that explains both baryogenesis and DM, with a potential connection to inflation. Nuggets of quarks formed in first-order phase transitions result in strangelets that are viable DM candidates~\cite{Carlson:2015daa}. 
These DM candidates are best probed through their interactions with nucleons, using large detectors. Current and proposed detection techniques range from repurposing current large-volume liquid noble detectors to specialized impedance-matched mechanical sensors \cite{CF1WP8}.
 
``Light" DM candidates which were in thermal equilibrium with the SM in the early universe inevitably lead to the notion of a hidden sector in which the DM interaction with the visible sector is mediated through a ``portal"~\cite{Battaglieri:2017aum,Essig:2022yzw}. Some of the most predictive of these models involve number-changing self-interactions of the DM, examples of which are the ``Elastically Decoupling Relic" (ELDER)~\cite{Kuflik:2015isi,Kuflik:2017iqs}, the ``Strongly-Interacting Massive Particle" (SIMP)~\cite{Hochberg:2014dra,Hochberg:2014kqa}, and DM from exponential growth~\cite{Bringmann:2021tjr}. One can also relax the requirement of thermal equilibrium in the early universe, and instead, populate the relic abundance through ``freeze-in"~\cite{Hall:2009bx}. Models of freeze-in require extraordinarily small couplings between the DM and the visible sector. Despite this, the presence of an ultralight mediator ($m_{\rm med}\ll$ keV) leads to a large enhancement of the direct-detection cross section at low momentum transfers. Importantly, such a model cannot be tested by accelerator-based probes; direct detection and cosmological probes provide the only possibility to test the freeze-in benchmark model.   Probes for sub-GeV DM are sensitive to the DM coupling to both electrons and nucleons, which provide an important handle for model discrimination. 

 \begin{figure}[t]
\begin{center}
\includegraphics[width=0.7\columnwidth]{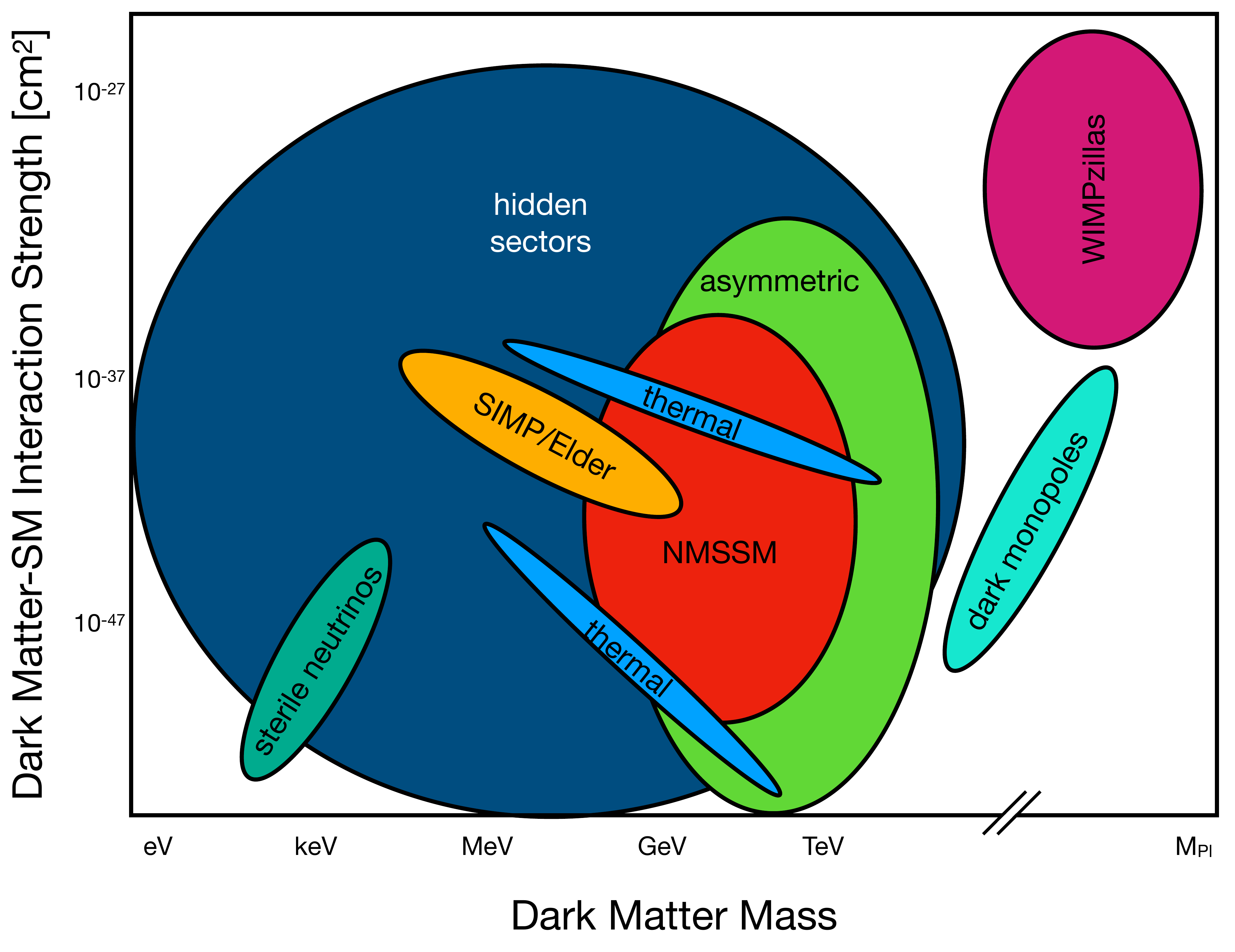}
\caption{Cartoon figure of the model space for direct detection. Included are candidates of thermal dark matter, supersymmetry, asymmetric dark matter~\cite{Kaplan:2009ag}, SIMP/Elder~\cite{Hochberg:2014dra,Hochberg:2014kqa,Kuflik:2015isi,Kuflik:2017iqs}, dark monopoles~\cite{Bai:2020ttp}, WIMPzillas~\cite{Chung:1998zb}, and hidden sector dark matter~\cite{Battaglieri:2017aum}. Note that the interaction cross-section can be for either scattering with nucleons or electrons, depending on the specific model.}
\label{fig:landscape_theory}
\end{center}
\end{figure}

\subsection{The path toward DM discovery with direct detection}
Many candidates in the ``heavy" range will not be tested by the suite of current generation experiments that are under construction or operating.  
The next suite of experiments should have an order of magnitude larger exposure and be able to significantly enhance our capabilities to probe much of this high-priority parameter space.
This future suite should probe models with spin-dependent interactions and others beyond the usual coherent DM-nucleus interactions. In addition, we cannot afford to eliminate support for successful DM search programs with unique sensitivity.  Similarly, many candidates in the ``light" range will not be tested with the current suite of ``small scale projects".  Continued investment to scale up in mass and/or reduce and understand low-energy backgrounds in programs to search for particle DM is thus crucial.

The benchmark for future generation experiments is to search for heavy DM candidates in the parameter space that reaches to the neutrino ``fog", the expected background from the coherent elastic neutrino-nucleus scattering (CE$\nu$NS) of solar and atmospheric neutrinos, or that advances sensitivity by an order of magnitude beyond the reach of current generation experiments in spaces where the fog remains many orders of magnitude distant, such as spin-dependent interactions.  For light mass DM candidates the goal over the next decade is to probe DM scattering down to 1 MeV and DM absorption down to 1 eV.
\subsubsection{Enabling Discovery with Complementary Probes}
\label{sec:DDsignatures}
The three categories of particle DM, as well as models within each category, give rise to distinct DM-SM interactions and experimental signatures in direct detection setups. Discovering particle DM requires a multi-faceted approach involving detectors that can measure different aspects of DM-SM interactions, as well as provide information about the DM distribution in our galactic halo.  

\begin{figure}
    \centering
\includegraphics[width=0.8\textwidth]{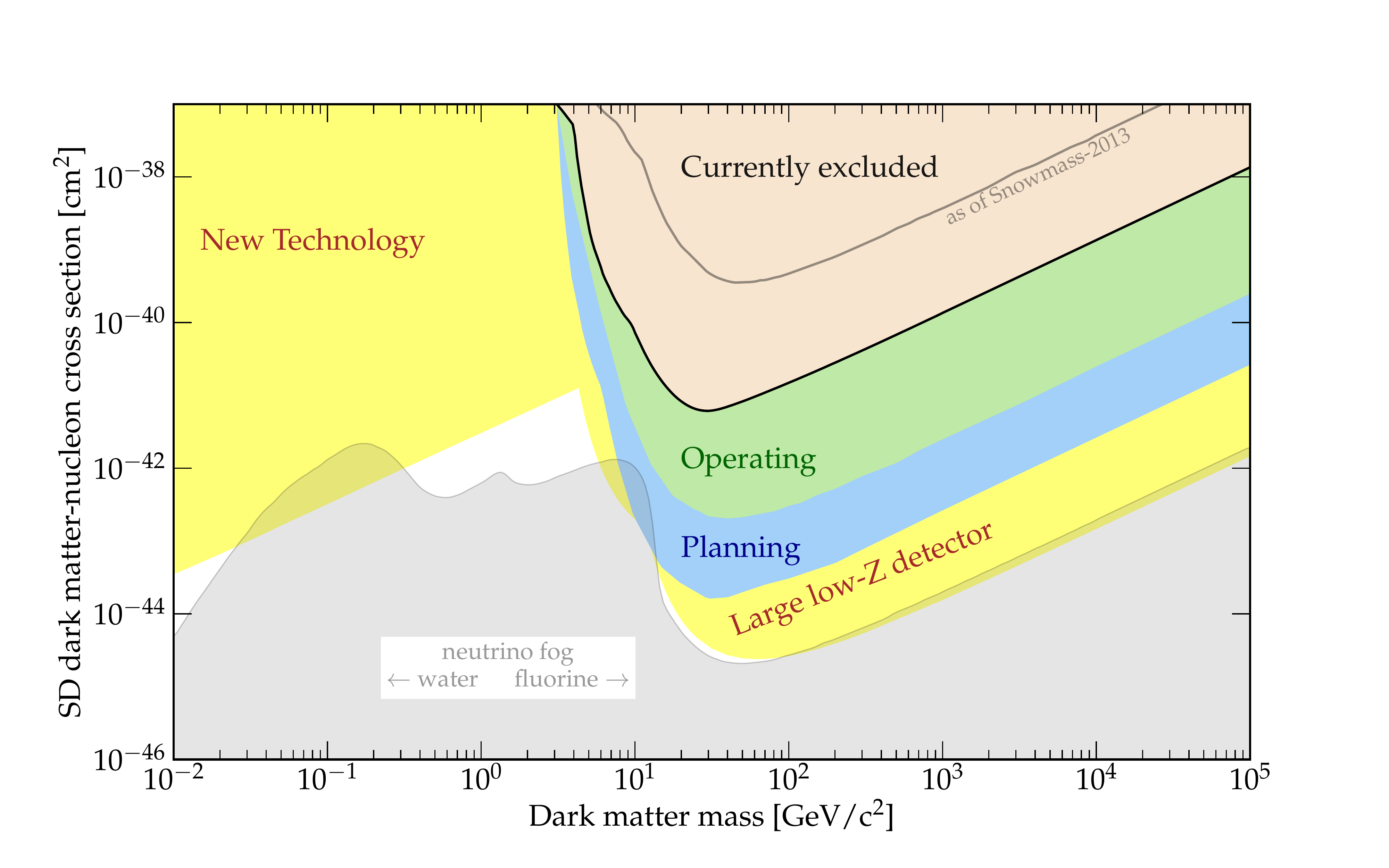}
   \caption{Combined Spin-dependent dark-matter nucleon scattering cross section space for scattering with neutrons or protons.
   Currently 90\% c.l.~constraints are shown in shaded beige ~\cite{Behnke:2012ys,Kim:2018wcl,Archambault:2012pm,Amole:2016pye,Amole:2017dex,CRESST:2019jnq,Archambault:2012pm,CDMS-II:2009ktb,LUX:2017ree,PandaX-II:2018woa,XENON100:2016sjq,Aprile:2019jmx,XENON:2019rxp} (data points taken from~\cite{OHare:2021utq}) 
   while the reach of currently operating experiments are shown in green (LZ, XENONnT). The limits from 2013 are shown as a gray line~\cite{Cushman:2013zza}. Future experiments are shown in blue (PICO-500, 1000 ton year Xenon) and yellow (Snowball, PICO-100 ton). The neutrino fog for a water or fluorine target is presented in gray, as described in Sec.~\ref{sec:neutrinofloor}. Plot reproduced from Ref.~\cite{CF1WP1}.}
    \label{fig:limitplot_SD}
\end{figure}

Heavy DM candidates, such as WIMPs, are traditionally probed via their interactions with {\bf nucleons} in the target material. Spin-independent interactions benefit from targets with high atomic mass due to the coherent $A^2$ enhancement of the scattering rate. On the other hand, spin-dependent interactions require a net spin of the nucleus, and hence an unpaired nucleon. At the same time, lighter DM candidates are most effectively probed by targets with lower $A$, where higher energy recoils are kinematically allowed and compensate for the lower DM kinetic energy. The current status and future prospects for spin-independent and spin-dependent interactions are illustrated in Figs.~\ref{fig:limitplot_SI} and \ref{fig:limitplot_SD}, respectively.

\begin{figure}
    \centering
    \includegraphics[width=0.45\textwidth]{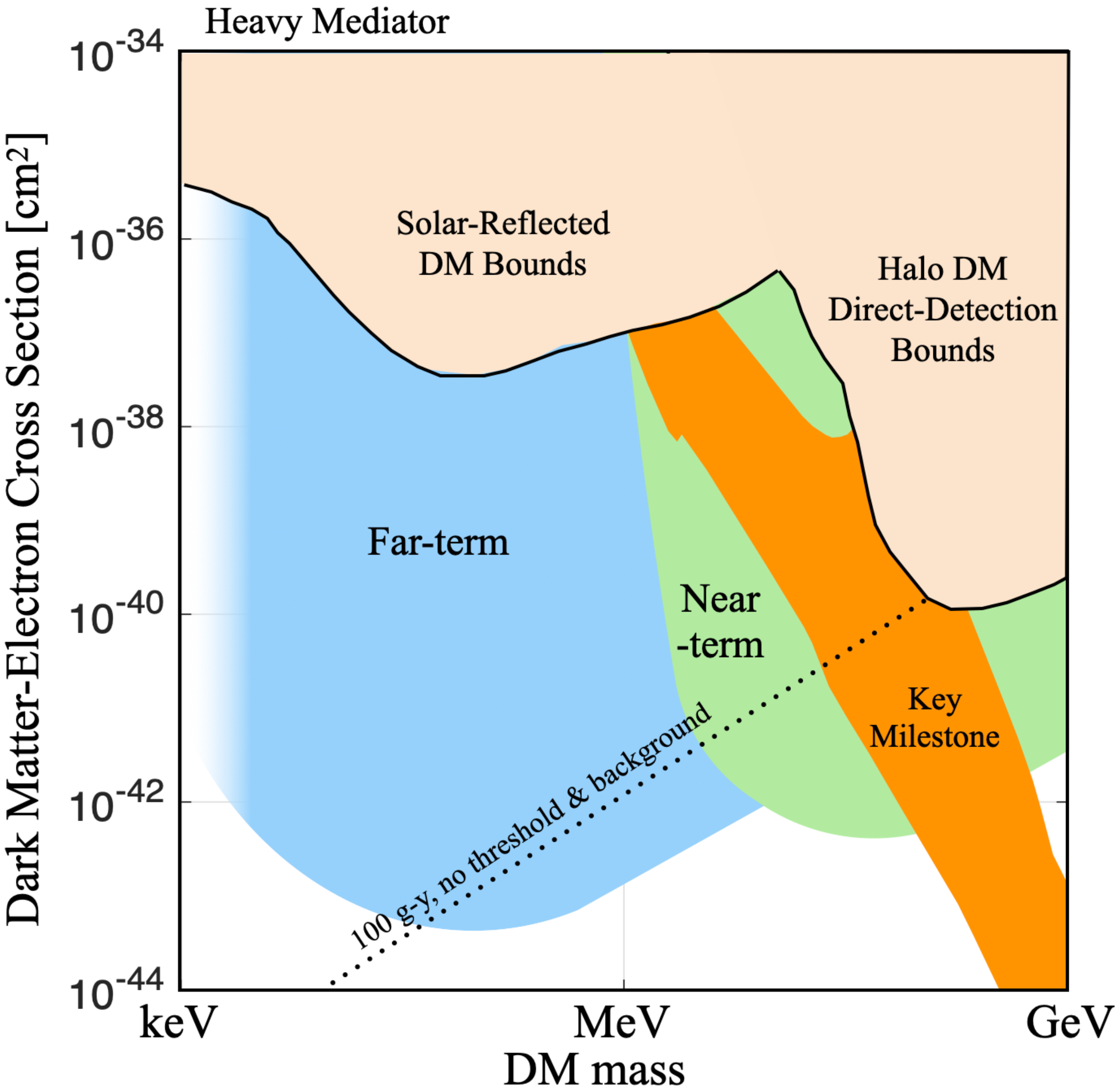}
    \includegraphics[width=0.45\textwidth]{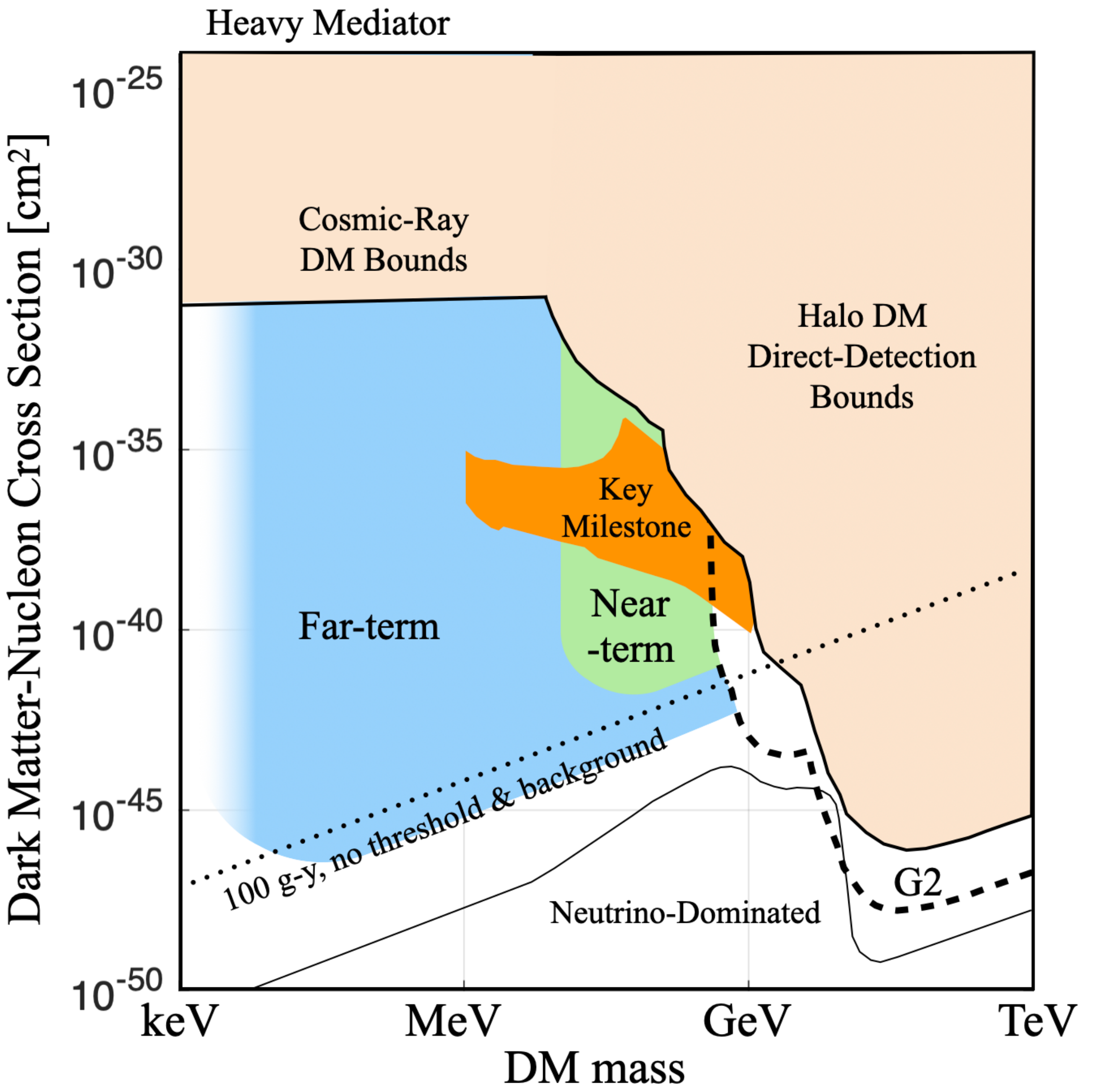}
    \includegraphics[width=0.45\textwidth]{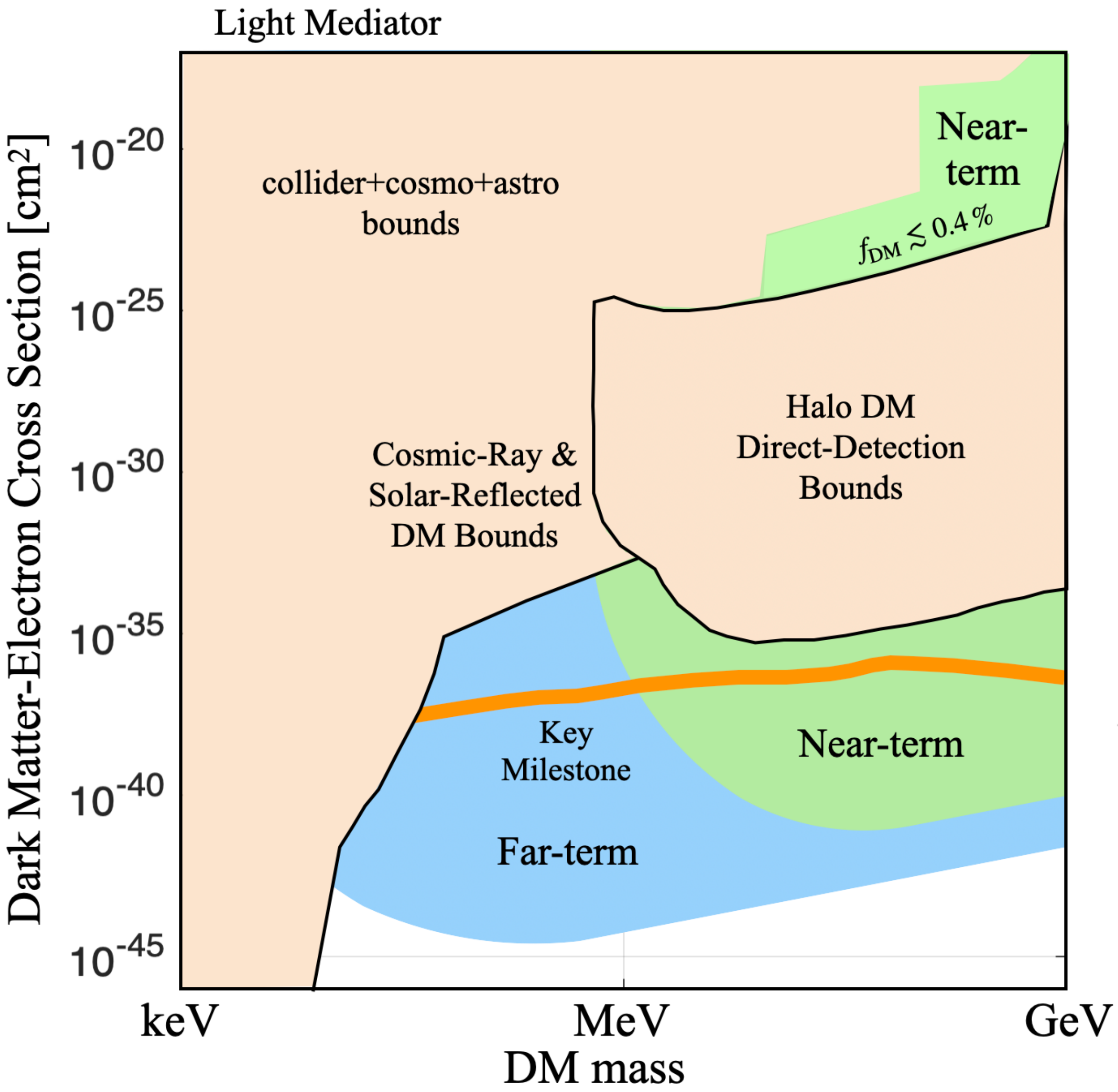}
    \includegraphics[width=0.45\textwidth]{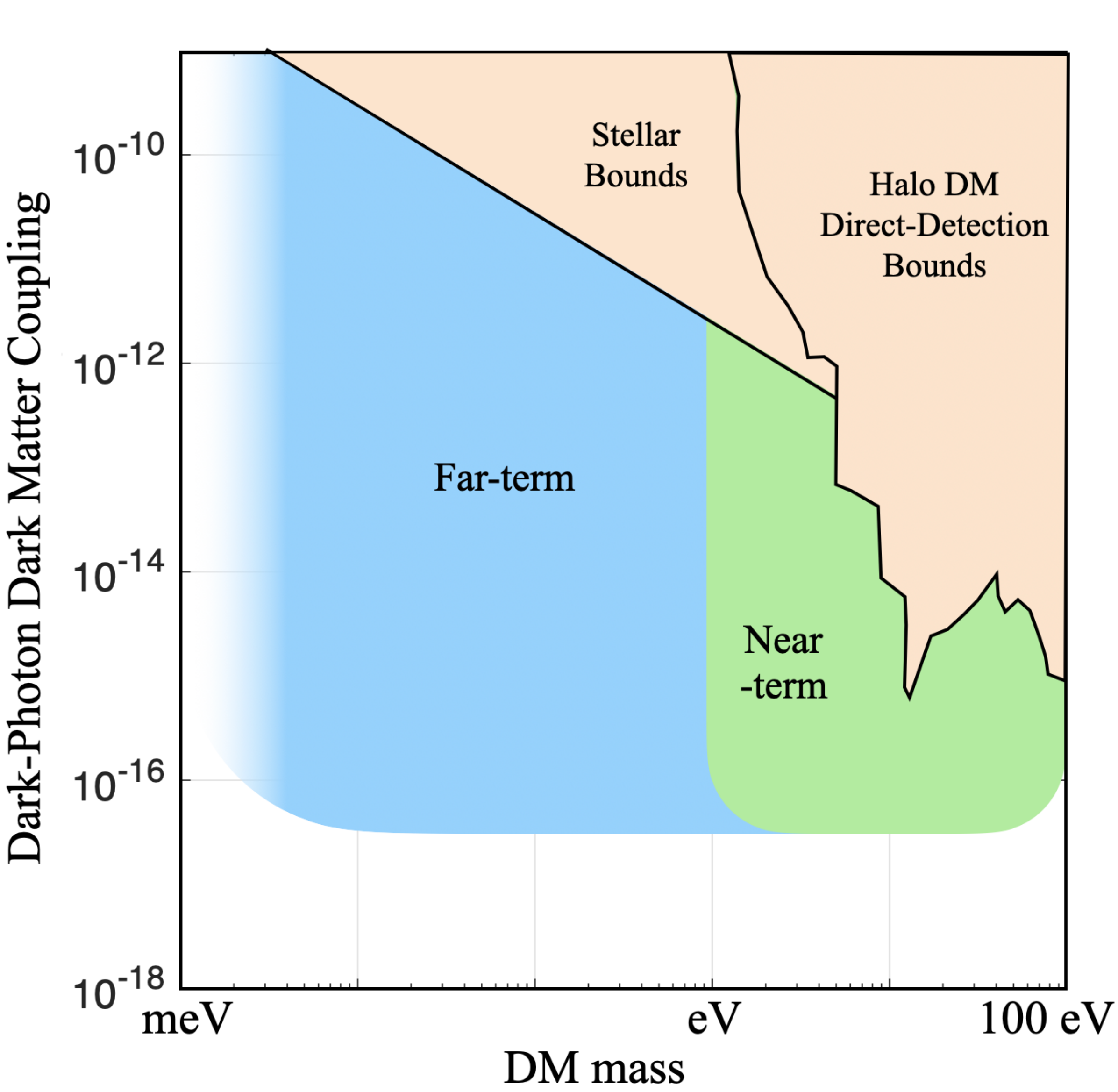}
    \caption{Figures are from Ref.~\cite{CF1WP2} and updated from BRN report~\cite{BRNreport}. Current 90\% c.l.~constraints are shown in beige. Approximate regions in parameter space that can be explored in the next $\sim$5 years (``near-term'', green) and on longer timescales (``far-term'', blue). Orange regions labelled “Key Milestone” represent concrete dark-matter benchmark models and are the same as in the BRN report~\cite{BRNreport}. Along the dotted line DM would produce about three events in an exposure of 100 gram-year, assuming scattering off electrons in a hypothetical target material with zero threshold.     
}
\label{fig:LDM_future}
\end{figure}

 In comparison, the direct detection of light DM was first probed through interactions with {\bf electrons} through DM-electron scattering as a result of the kinematic matching of the DM and electron mass (bosonic candidates were also detected through absorption)~\cite{CF1WP2}. These interactions can be detected via ionization, scintillation, or phonon signatures depending on the experimental set-up. In addition, light DM can also be probed via DM-nucleon scattering through either a bremsstrahlung process~\cite{Kouvaris:2016afs} or what is known as the ``Migdal effect"~\cite{Ibe:2017yqa}; both of these rely on detecting secondary photons or electrons that are a consequence of the primary DM-nucleon interaction. 
 One of the benchmark models for light DM candidates is a dark photon mediated sub-GeV DM candidate. In this scenario, the DM will interact with {\bf both} electrons and charged nucleons. In contrast, a leptophilic mediator will lead solely to DM-electron interactions while a leptophobic mediator will only have DM-nucleon interactions. 
 Thus, it is imperative to calculate both DM-electron and DM-nucleon scattering rates, comparisons of which will provide further handles on the underlying DM model governing the interactions. For DM candidates below the MeV mass scale, DM interactions with collective modes in the material such as {\bf phonons} are key to discovery due to the low energy-thresholds. The landscape of sub-GeV DM direct detection, including current constraints and future projections, is illustrated in Fig.~\ref{fig:LDM_future}. Notably, for sufficiently large nuclear scattering cross sections even (very) light DM particles can be robustly probed by conventional direct detection experiments~\cite{Bringmann:2018cvk}, because cosmic rays inevitably up-scatter a sub-population of such particles to relativistic energies (thus allowing for sufficiently high recoil energies).

\begin{figure}[t]
\centering
\includegraphics[width=1\linewidth]{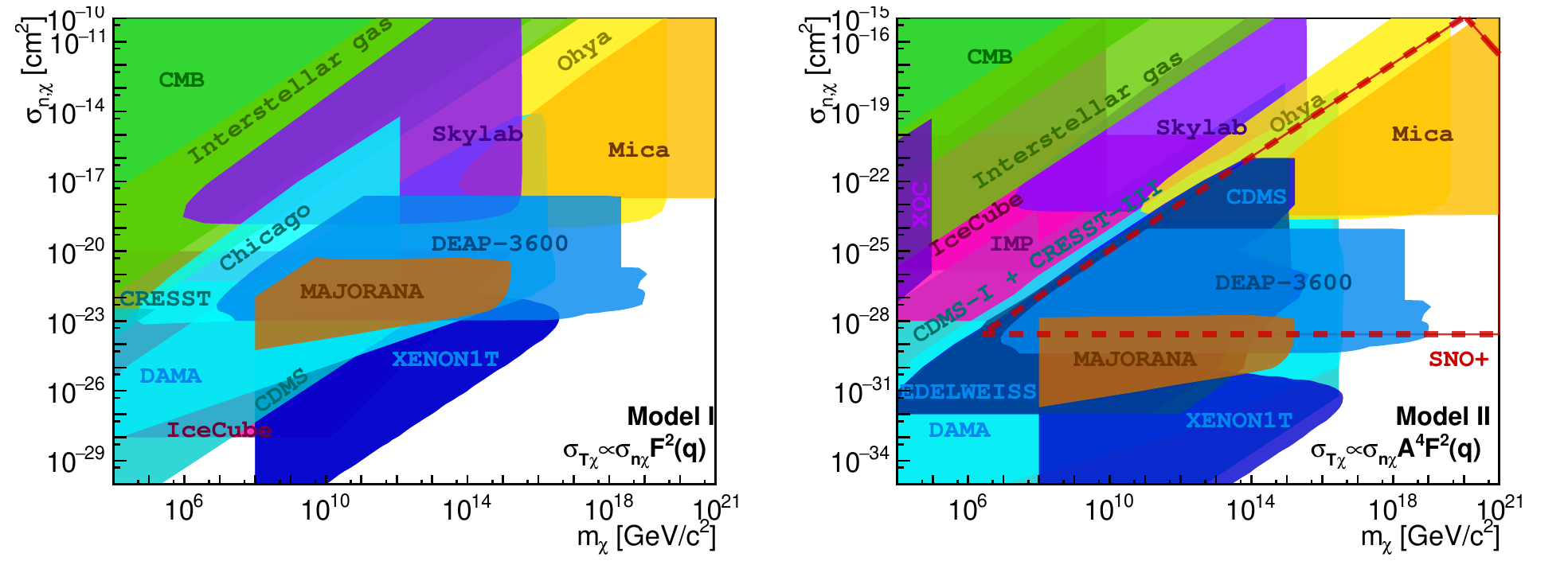}
\caption{Current and projected experimental regions of ultraheavy parameter space excluded by cosmological/astrophysical constraints (green), direct detection DM detectors (blue), neutrino experiments (red/orange), space-based experiments (purple), and terrestrial track-based observations (yellow). Both models considered here assume different relations for the cross section scaling from a single nucleon to a nucleus with mass number $A$. In the left plot, we assume no scaling with $A$; in the right plot, we assume the cross section scales like $A^4$ (\eg, with two powers coming from nuclear coherence, and two from kinematic factors). Limits are shown from DEAP-3600~\cite{DEAPCollaboration:2021raj}, DAMA~\cite{bernabei_extended_1999,BhoonahSkyLab:2020fys}, interstellar gas clouds~\cite{BhoonahGasClouds:2018gjb,BhoonahGasClouds:2020dzs}, a recast of CRESST and CDMS-I~\cite{kavanagh_earth-scattering_2018}, a recast of CDMS and EDELWEISS~\cite{albuquerqueDirectDetectionConstraints2003,albuquerqueErratumDirectDetection2003}, a detector in U.~Chicago~\cite{cappiello_new_2021}, a XENON1T single-scatter analysis~\cite{Clark:2020mna}, tracks in the Skylab and Ohya plastic etch detectors~\cite{BhoonahSkyLab:2020fys}, in ancient mica~\cite{AcevedoMica:2021tbl}, the MAJORANA demonstrator~\cite{Clark:2020mna}, IceCube with 22 strings~\cite{albuquerqueClosingWindowStrongly2010}, XQC~\cite{erickcekConstraintsInteractionsDark2007}, CMB measurements~\cite{Dvorkin:2013cea,Gluscevic:2017ywp}, and IMP~\cite{IMP}. Also shown is the future reach of the liquid scintillator detector SNO+ as estimated in \cite{Bramante:2018tos,Bramante:2019yss}. Reproduced from Ref.~\cite{CF1WP8}.}
\label{fig:UHDM_dd_limits}
\end{figure}

UHDM faces a unique challenge in that the flux of DM is quite low, $\Phi\simeq 0.85 {~\rm events}/({\rm m}^2{\rm yr})\times (M_{\rm Pl}/m_\chi)$. However, the signatures of UHDM are striking. 
The kinematics of UHDM are relatively unaffected by the scattering with nuclei in the target material and the UHDM effectively travels undeflected through the detector. As a result, the UHDM may scatter on multiple nuclei as it crosses the detector, leaving behind a track that traces the DM velocity vector~\cite{Bramante:2019yss}. UHDM can also be probed via indirect detection through neutrino observatories. Fig.~\ref{fig:UHDM_dd_limits} displays the complementarity of UHDM direct detection probes.

For all classes of models, the scattering rate, and therefore sensitivity of the detector, scales with the detector exposure, (fiducial target mass)$\times$(exposure time). One way to increase the exposure is by increasing the fiducial mass of the target. However, another option is to increase the exposure time, \eg proposed searches using paleodetectors~\cite{Baum:2018tfw,Drukier:2018pdy,Edwards:2018hcf} have exposure times of order 100 Myr. Interestingly, the multiple scatter signatures of UHDM imply that the {\it area} of the detector is the most relevant factor rather than the {\it volume} of the detector~\cite{CF1WP8}. 

As DM detectors increase in sensitivity, they begin to probe the ``neutrino fog" (described in Sec.~\ref{sec:neutrinofloor}) and strategies to distinguish between neutrino and DM signatures become necessary. Some mitigation strategies include the use of annual modulation of the DM rate~\cite{Drukier:1986tm} and directional detectors (see \eg review found in~\cite{Vahsen:2021gnb}). Even more tantalizing, directional detectors have the potential to map out the DM substructure in our galaxy.   

\subsubsection{Navigating the Neutrino Fog}
\label{sec:neutrinofloor}
{\it This section draws heavily from WP1, 
Section 2~\cite{CF1WP1}}

As direct DM searches accumulate ever larger exposures in pursuit of testing smaller interaction cross sections, the expected background from the coherent elastic neutrino-nucleus scattering (CE$\nu$NS) of solar and atmospheric neutrinos increases. For DM masses $\gtrsim1$~GeV/$c^2$, the relevant neutrino fluxes are those from $^8$B solar neutrinos, atmospheric neutrinos, and the diffuse supernova neutrino background (DSNB). The CE$\nu$NS event rate from each of these is subject to a systematic uncertainty which manifests from the uncertainty on the flux of each neutrino species. In addition, the nuclear recoil energy spectra of these events closely resembles that of the sought after DM signal. 

The effect of these backgrounds -- particularly because of their associated systematic uncertainties and spectral shapes --  will be to reduce the DM sensitivity achievable as the experimental exposure grows. This fact led to the creation of the colloquially known ``neutrino floor": a boundary in the cross section versus DM matter mass plane below which a DM discovery becomes extremely challenging without further constraints on the neutrino backgrounds. The precise location of this boundary has evolved over the last decade~\cite{Billard:2013qya,Gelmini:2018ogy,OHare:2020lva,OHare:2021utq}, and more recent studies~\cite{OHare:2020lva,OHare:2021utq} have emphasized a crucial point: that astrophysical neutrino backgrounds do not impose a hard limit on physics reach; rather, the effect is more gradual than a single boundary depicts. To reinforce this concept within the community, we now adopt the term {\bf neutrino fog} to describe the region of DM cross sections where neutrino backgrounds begin to inhibit the progress of direct detection searches.

The neutrino fog is quantified by the index $n$, defined as the gradient of a hypothetical experiment's median cross section for $3\sigma$ discovery with respect to the exposure:
\begin{equation}
    n = -\bigg( \frac{d \log{\sigma}}{d \log{MT}} \bigg)^{-1}
\end{equation}

This index quantifies the diminishing return-on-investment in increasing exposure when limited by neutrinos. Specifically, if an experiment has achieved some cross-section sensitivity, further reducing the sensitivity by a factor of $x$ requires increasing the exposure by at least $x^n$. \eg, at $n=2$, reducing the cross section reach by a further factor of 10 requires increasing the exposure by a factor of at least 100. 

\begin{figure}
    \centering
    \includegraphics[width=0.8\textwidth]{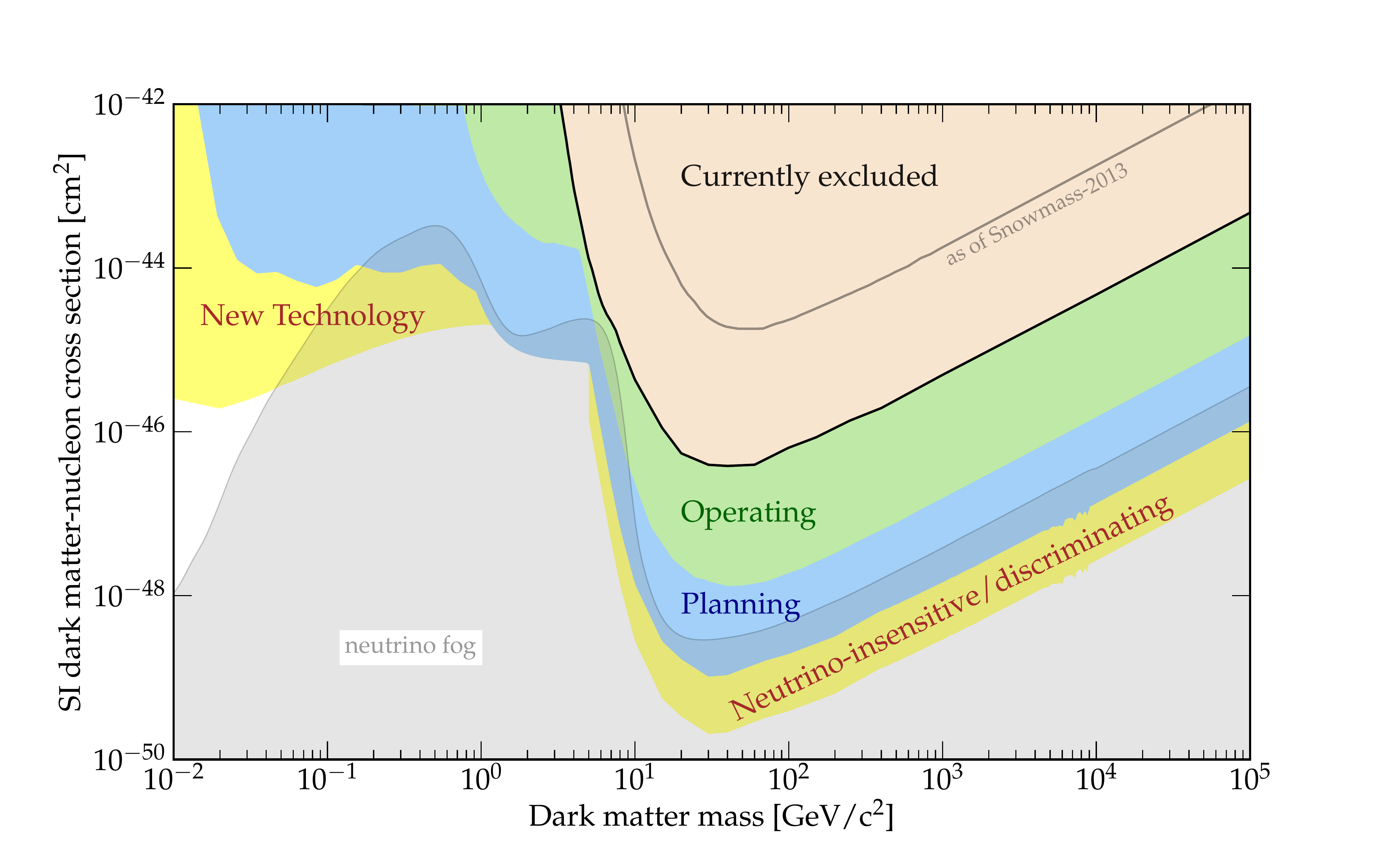}
   \caption{Combined Spin-independent dark-matter nucleon scattering cross section space. 
   Currently 90\% c.l.~constraints are shown in shaded beige
   ~\cite{XENON:2020gfr,SuperCDMS:2018gro,Adhikari:2018ljm,CRESST:2019jnq,Bernabei:2018yyw,DarkSide:2018bpj,DEAP:2019yzn,EDELWEISS:2016nzl,LUX:2016ggv,NEWS-G:2017pxg,PandaX-II:2017hlx,PICO:2016kso,PICO:2017tgi,XENON:2019zpr,PandaX-4T:2021bab,LZ:2022ufs} (data points taken from~\cite{OHare:2021utq})
   while the reach of currently operating experiments are shown in green (LZ, XENONnT, PandaX-4T, SuperCDMS SNOLAB, SBC). The limits from 2013 are shown as a gray line~\cite{Cushman:2013zza}. Future experiments are shown in blue (SuperCDMS, DarkSide-LowMass, SBC, 1000 ton-year liquid xenon, ARGO) and yellow (Snowball and Planned$\times$ 5). The neutrino fog for a xenon target is presented in gray as described in Sec.~\ref{sec:neutrinofloor}. 
   Plot reproduced from Ref.~\cite{CF1WP1}.}
    \label{fig:limitplot_SI}
\end{figure}

In Figure~\ref{fig:limitplot_SI}, the neutrino fog is shown for the index $n=2$, which marks the transition from statistically- to systematically-limited.
A critical feature of the neutrino fog is that it will move to lower cross section if uncertainties in the neutrino fluxes are reduced, opening up new space for continuing searches. Therefore, all future detectors entering into the neutrino fog would benefit from improved atmospheric neutrino background modeling, which currently dominates the uncertainty on the experimental sensitivity. 


\subsubsection{Tools to resolve signals from backgrounds }
{\it This section draws heavily on WP3~\cite{CF1WP3}.}

The next generation of direct detection DM experiments will require more precise modeling of signal and background rates.  
This goes hand-in-hand with calibration measurements of detector responses to nuclear and electronic recoils for a wider range of targets and at lower energy scales, including inelastic and atomic effects and coherent excitations.

\begin{itemize}
    \item Optical and atomic material-property measurements are needed, including atomic de-excitation cascades, electronic energy levels, and energy-loss functions, both for modeling detector response and for simulating transport of low-energy particles. Similarly, improved models and simulations of detectable-quanta production and propagation are needed.
    \item Calibration measurements are needed of detector responses to electronic and nuclear recoils, for a wider range of targets and at lower energies  (eV--keV scale and in the ``UV-gap'' of solid-state detectors), including measurements of inelastic and atomic effects (\eg\ the Migdal effect) and coherent excitations, both using established techniques and with new ones
    \item  Measurements to decrease neutrino uncertainties are needed, including nuclear-reaction and direct-neutrino measurements that improve simulation-driven flux model.
    \item Nuclear reaction measurements—($\alpha$, n), (n, $\gamma$), (n, n), ($\nu$,x), etc.—are needed to improve particle transport codes, background simulations, and material activation calculations, and to validate and improve models/evaluations used in these codes; exclusive cross-section measurements are particularly needed to model correlated ejectiles, and codes should provide a full treatment of uncertainties.
    \item Increased collaboration with the nuclear, atomic, condensed matter, and cosmic-ray physics communities is recommended to improve detector response and background models.
    \item In situ measurements of backgrounds are valuable for validating and improving models in various codes, especially for neutrons and muons and material activation measurements, both on surface and underground. Likewise, uncertainties on model predictions for these backgrounds need to be quantified by model codes, informed by this validation.  
    \item Improved material cleaning and screening procedures are needed for dust, radon progeny, cosmogenically activated radioisotopes, and other bulk radioisotopes. New procedures are needed to avoid such contamination, and improved ex situ models of residual background levels are needed.
    \item Additional R\&D is needed to model, measure, and mitigate near-threshold backgrounds, such as from secondary emission processes.
    
    \item Same-location, multi-method radioassay facilities are needed to simplify measurements of decay chains and reduce systematic errors, alongside greater precision across techniques and increased assay throughput/sensitivity. Also, development of software infrastructure to track large-scale assay programs across the community is needed.
    \item Measurements of the ionization topology of nuclear recoils can be performed with recoil imaging directional detectors \cite{OHare:2022jnx}. This will provide improved validation of nuclear recoil simulation tools such as SRIM, and first experimental confirmation of the neutral Migdal effect \cite{Vahsen:2021gnb, Araujo:2022wjh}.
    \item Finally, investment in underground infrastructure and detector technology (\eg vetoes) will be needed to mitigate backgrounds such as from radon, dust, cosmogenic activation, and neutrons.
\end{itemize}

\subsubsection{Simulations and Computing to enable particle dark matter detection}
{\it This section draws heavily on WP4~\cite{CF1WP4}.}

Simulations are an essential component of modern physics experiments to translate between abstract physics models and a real implementation of a detector. They are vital at every stage of the detector life-cycle, starting with the initial design phase, all the way to final data analysis. In some cases, the associated costs to support computing and code development may represent a large fraction of the experiment's budget. Therefore, it is essential to have an efficient, well-maintained, well-understood and thoroughly validated simulation infrastructure. A full-chain simulation infrastructure includes both event generators and detector simulation frameworks. Therefore, the direct detection DM community has a clear need for the following:
\begin{itemize}
    \item Increased support for development of simulations (\eg\ \fluka, \geant, \mcnp, \nest, \gfcmp), both for software codes and models. This includes support and training within the community, and increased opportunities for cross-collaboration communication in order to reduce duplication and maximize return on investment. 
    
\end{itemize}

Current DM experiments are approaching data volumes of order 1 PB/year~\cite{Mount:2017qzi}. This is not an unusual scale for HEP, however it presents a significant challenge in the direct detection community, which until recently has not prioritized the development of a scalable computing infrastructure in support of its scientific ambition. Moreover, a fragmentation of funding sources and a climate of competition between experiments, have hindered the opportunities for cooperation and tool sharing in the computing and software domain, leading to unnecessary duplication of efforts across the field. Key strategic goals to ensure success for the experiments of the next decade include:

\begin{itemize}
    \item Lower the barrier of entry to national supercomputing facilities, by providing common tools and shared engineering. Solicit community input on architecture evolution, while providing access to specialized resources, such as GPU and TPU clusters.
    \item Support scalable software infrastructure tools across HEP, avoiding duplication of effort. These tools run the gamut of data management and archiving, event processing, reconstruction and analysis, software management, validation and distribution.
    \item Enhance industry collaborations on machine learning techniques and provide access to external experts. Foster community-wide efforts to understand uncertainties and physical interpretation of machine learning results.
\end{itemize}

\subsubsection{Resolving excesses and making a discovery}
{\it This section draws heavily on WP6~\cite{CF1WP6}.}

As with indirect detection, excesses have also been reported and disputed within the direct detection community. A strength of direct detection is that experiments can be replicated or altered in response to such excesses. While not a trivial exercise, replication is an option of last resort to fully confirm or rule out any potential signal for dark matter. We take some recent examples as case studies. 
The DArk MAtter/Large sodium Iodide Bulk for RAre processes (DAMA/LIBRA) collaboration operates ~250 kg  ultra-low background NaI scintillators as dark matter detectors at the Gran Sasso National Laboratory (LNGS), Italy. It has consistently reported an excess of modulating low energy events between 2-6 keV$_{ee}$ region as the annual modulation signal of dark matter~\cite{DAMAPhase2}, consistent with expected modulation signal from dark matter particle interactions. The collaboration claims that no systematics or side reaction can mimic the annual modulation signal~\cite{DAMABkg}. The DAMA/LIBRA claimed dark matter signal is controversial because other direct detection experiments with different target materials (Xe, Si, Ge) have failed to observe the signal based on standard WIMP models. To fully resolve this dilemma, independent experiments with the same target material (NaI(Tl)) as DAMA have been carried out or are proposed~\cite{ANAIS-AM, ANAIS-AM2, ANAIS-AM3, COSINE-AM, COSINE-AM2,COSINE200, SABARE, PICOLON, COSINUS-2020}. In particular, the COSINE-100 collaboration and the Annual modulation with NaI Scintillators (ANAIS) collaboration have made good progress on probing the DAMA signal with the same target material. ANAIS-112 has refuted the DAMA/LIBRA positive modulation with almost 3$\sigma$ sensitivity in a model independent way with three years of data, and is expected to surpass 4$\sigma$ sensitivity after completing six years of data taking (along with 2023 data), while COSINE-100 has excluded WIMPs as responsible of the DAMA/LIBRA signal in many scenarios. 
The two experiments have initiated discussions to conduct a combined analysis to search for annual modulation signals, and a new experiment SABRE~\cite{Froborg:2016ova,SABRE:2018lfp}, with installations in both northern and southern hemispheres, is preparing to run in several years. 

Another path for resolving the nature of the excess (or any similar future excess) is by examining the directionality of the signal. Unlike the annual rate and energy modulation, the sidereal-daily directional modulation (equivalent to a dipole in galactic coordinates), is expected to be immune against time-dependent backgrounds originating in the laboratory. Observing this directional signal would therefore provide a robust confirmation that a tentative excess is real and of galactic origin, but improved detector capabilities are required (see Section~\ref{sec:directionality} \cite{Vahsen:2021gnb}).

More recently, a low-energy electronic recoil (ER) excess below 7\,keV and most prominent between 2--3\,keV was observed in the XENON1T dark matter experiment in 2020~\cite{XENON:2020rca}. With a significance of $\sim3.5 \sigma$, this excess could have been a statistical fluctuation, a hint of a new background process, or a hint of new physics. In this case, the natural progression of direct detection enabled the successor experiment, XENONnT, to directly test this signal. With just a few months of data, XENONnT was able to rule out this signal, attributing the previous result to an unmodeled background~\cite{XnT2022}. The LZ experiment will also have complementary sensitivity in the near future.

Several solid state experiments have also seen excesses at low-energies, which has led to the EXCESS series of workshops~\cite{Proceedings:2022hmu}. These workshops are a community-wide effort which finds competing experiments, using different solid state technologies, collaborating together to characterize and resolve the excesses. These efforts may lead to shared technologies and combined R\&D efforts, which are crucial for further progress in DM research. 

\subsection{Experiments and R\&D}
{\it This section draws heavily on WP1 and WP2~\cite{CF1WP1,CF1WP2}, as well as~\cite{OHare:2020lva}}.

Direct detection experiments can be broadly classified based on the DM mass range they are particularly targeting or based on the technology used in the experiment, although many do cross boundaries in both categories. These experiments tend to follow a consistent life cycle - an early, demonstration phase where sensitivity is proven followed by a scaling phase where the technology is optimized and the target mass is increased to improve sensitivity. New backgrounds are often discovered with each new iteration, as old backgrounds are understood and mitigated. 

Until $\sim$ 2010, most direct detection efforts focused on the ``traditional'' WIMP mass range of $\gtrsim$1 GeV; the current generation experiments, which are the modern successors to those experiments, are well into the scaling phase, featuring mature technologies with large target masses. The goal of these current collaborations is to continue scaling up to even larger targets to push the sensitivity down to the neutrino fog. In the last decade, a compelling case for light DM particles has been made from a theoretical standpoint and new techniques have been developed to match. These experiments that target light mass DM tend to be much smaller in scale with a focus on achieving very low energy thresholds, and they are now proving that they can have excellent sensitivity to previously unexplored regions of DM parameter space. The understanding of new backgrounds that have shown up at the lowest energy thresholds in these detectors is only now beginning for these detectors which are not fully in the scaling phase.  Table~\ref{table:DD} provides a tabular overview of the field of direct detection. Ref.~\cite{CF1WP1} provides an excellent summary of the direct detection efforts aimed at DM masses $\gtrsim$1 GeV, and Ref.~\cite{CF1WP2} does the same for DM masses $\lesssim$1 GeV. We will only briefly re-summarize those reviews here and suggest the reader refer to the full length reports and references therein for more information. 

The rough categorization of direct detection techniques presented below is not intended to be exclusive. In addition, many of the techniques discussed span multiple categories. Solving the mystery of dark matter will take all the ingenuity that physicists can bring to bear, and supporting new ideas is a key component of a successful program. 

\subsubsection{Dark Matter Candidates with Mass Greater Than 10 GeV}
For experiments seeking to probe down to the neutrino fog, it is helpful to further subdivide the DM mass space into $<$ 1-10 GeV and greater than 10 GeV. 
The latter category includes the canonical WIMP, and represents the most explored region of DM phase space by direct detection experiments. In this region, liquid xenon (LXe), liquid argon (LAr), and freon-based bubble chambers are the dominant technologies. 

\begin{figure}[!htbp]
\begin{center}
\includegraphics[width=0.68\columnwidth]{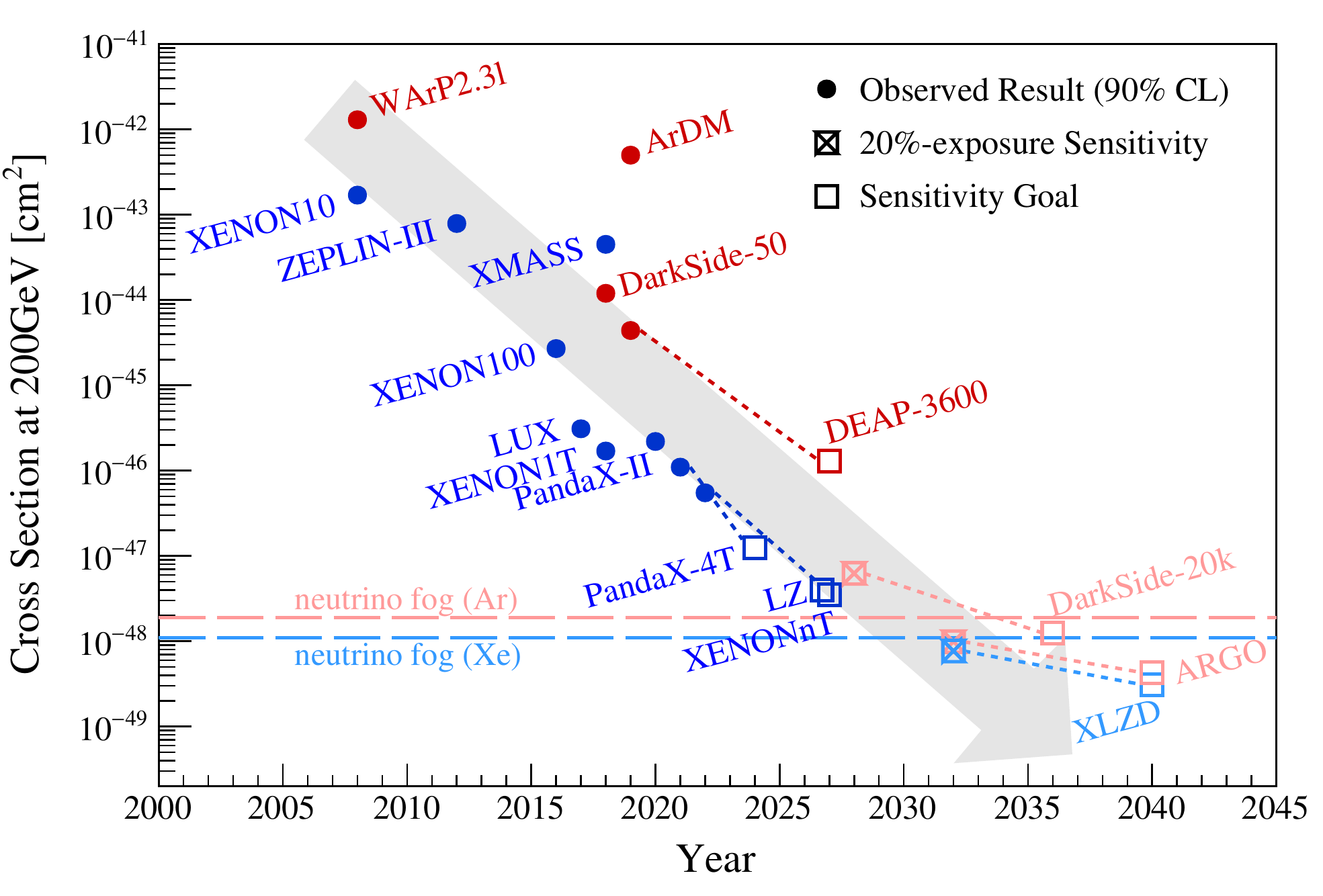}
\caption{Figure adapted from~\cite{Aalbers:2022dzr}. Improvement in sensitivity to spin independent WIMP-nucleon coupling for a DM mass of 200~GeV/c$^2$ achieved by LXe (blue) and LAr (red) experiments of increasing target masses with sensitivities (squares) and the best achieved limits
(filled circles), for both past (dark color) and future (light color). Open squares correspond to the sensitivity at the goal exposure while the crossed squares correspond to the sensitivity at 20\% of goal exposure. Cross section values and background rates are extracted from Ref.~\cite{PhysRevLett.100.021303,PhysRevD.94.122001,AKIMOV201214,PhysRevLett.116.161301,Wang_2020,XENON:2018voc,PandaX:2018wtu,PandaX-4T:2021bab,AKERIB201804,XENON:2020kmp,Fatemighomi:2016ree,DEAP:2019yzn,Wang:2017jts,20.500.11850/335024,BENETTI2008495,WARP,Agnes:2018ep,Darkside20k,DarkSideARGO}.
}
\label{fig:xenon_evolution}
\end{center}
\end{figure}

\subsubsection*{Noble Element Detectors}

Figure~\ref{fig:xenon_evolution} summarizes the impressive achievements of the LXe-time projection chamber (TPC) technology over the last two decades,
with the LZ experiment presently holding the most stringent constraints for WIMP masses above 9\,GeV/c$^2$ (with PandaX-4T above 0.1\,GeV/c$^2$ if the Migdal effect is assumed)~\cite{Aalbers:2022dzr,PhysRevLett.123.251801, PhysRevLett.123.241803,PandaX-4T:2021bab}.
All three multi-tonne LXe-TPCs -- LZ, XENONnT, and PandaX-4T -- are now operating with a target sensitivity of about 1-2$\times$10$^{-48}$ cm$^2$ (see Fig.~\ref{fig:xenon_evolution}).

The DARWIN collaboration~\cite{DARWIN:2016hyl} has already started in Europe an intense R\&D program for a 40-80 tonne Xe target. Given their current leadership roles and expertise, the US teams in both XENONnT and LZ (about 50\% of the community) are well-positioned to contribute significantly to this next-generation effort.   
In 2021, scientists from the LZ and XENON/DARWIN collaborations joined in what is now known as the XLZD consortium~\cite{mou} towards jointly building and operating this next-generation xenon-based experiment.
There would be substantial opportunity cost in delaying the US engagement.
The sensitivity for an exposure of 1\,ktonne$\cdot$year delivering a $3\sigma$ DM particle discovery~\cite{Aalbers:2022dzr}
is illustrated in Fig.~\ref{fig:limitplot_SI}. 
The strategy to approach such an exposure will be informed by outcomes from the current projects, complemented by a well-defined set of R\&D activities that should take place concurrently. 
Community support for a next generation xenon-based experiment is very broad. A 600-author white paper outlining the physics reach of such an experiment was recently published~\cite{Aalbers:2022dzr}. In China, PandaX-4T
is expected to continue operations until 2025 and then to be incrementally upgraded into a multi-ten-ton experiment with a nominal fiducial target of 30~ton.

The large detector scale and low background, combined with many advantageous properties of xenon properties as target, enable a rich science portfolio that extends beyond WIMPs, transforming such a detector into a cost-effective, broad, low-background astroparticle physics observatory. Given the significant reach of multi-ten-tonne scale xenon detectors and the relatively inelastic market for procuring xenon in these quantities, R\&D is needed to support purpose-built and/or alternative procurement and refining capabilities. Reducing uncertainty regarding the time and cost to acquire xenon addresses a critical risk to advancing this well-proven leading technology.

An unambiguous positive direct detection of DM will require exceptionally strong evidence of a signal. The dual-phase TPC with underground argon (UAr) is a powerful emerging technology that potentially remains free of all backgrounds except for CEvNS at the largest detector scales. More than 400 scientists from over 60 institutions, including scientists from the ArDM, DarkSide-50, DEAP-3600, MiniCLEAN, and XENON collaborations, formed the multinational Global Argon Dark Matter Collaboration (GADMC). The GADMC, which has significant US involvement, is committed to pursuing the search for DM using an UAr target with the DarkSide-20k (DS-20k) and Argo experiments. DS-20k will build upon the successful elements of DarkSide-50 and DEAP-3600, including the use of UAr, which DarkSide-50 showed to have an $^{39}$Ar concentration at least 1400 times smaller than atmospheric argon~\cite{Agnes:2015gu,Agnes:2016fz,Agnes:2018ep}, and a radiopure acrylic structure, which is a technology pioneered by the DEAP-3600 experiment~\cite{Nantais:2013jp,Amaudruz:2018gr}. The TPC is designed to take advantage of the uniquely powerful properties of liquid argon, including excellent chemical purity~\cite{Ajaj:2019jx,Ajaj:2019jk} and demonstrated electron recoil (ER) background discrimination power of better than $10^8$~\cite{Adhikari:2021}. In addition, DEAP-3600 has demonstrated the lowest radon contamination for any large noble liquid detector to date, at 0.15 uBq/kg~\cite{Ajaj:2019jk}.

Unique among all DM and $0\nu\beta\beta$ experiments, the GADMC will produce its own low-radioactivity UAr target  ($^{40}$Ar, with $^{39}$Ar and $^{42}$Ar strongly suppressed). The UAr is extracted in the US and further chemically purified in Italy. The GADMC will produce several hundred tonnes of UAr without dependencies on uncertain/unstable international sources. In the same spirit, the GADMC developed its own Si-based photosensors (SiPMs) and constructed a special facility at LNGS, NOA, hosted in a radon-suppressed clean room, for the packaging of SiPMs into large-scale (several tens of $m^2$) planes of radio-pure, cryogenic photodetectors. These developments set the stage for Argo, a planned DM and neutrino observatory, which will reach deeply into the neutrino fog, as shown in Figure~\ref{fig:limitplot_SI}.

Work is underway to explore the feasibility of a low background kTon-scale liquid argon time projection chamber in the context of the Deep Underground Neutrino Experiment (DUNE).  The DUNE program consists of modules, and though the designs for the first two modules have been selected, the third and fourth (the `modules of opportunity’) remain to be determined.  A recent community effort~\cite{SnowmassLowBgDUNElikeWP} explored the option of making one of these a dedicated low background module, with possible sensitivity to high mass WIMPs~\cite{Church_2020}. 
The primary research and development challenges for this detector design are associated with the large scale, representing a further order of magnitude from the Argo concept. 

\subsubsection*{Phase change detectors}

Bubble chambers present a scalable, background-discriminating technology for DM detection with the unique capability to operate with a broad variety of target materials, requiring only that the target be a fluid with a vapor pressure. 
The PICO Collaboration uses freon-filled bubble chambers for nuclear recoil detection in targets with high spin-dependent and low spin-independent cross-sections.
With field-leading electron recoil rejection ($O$(1) in $10^{10}$ ER events misidentified as nuclear recoils~\cite{PICO:2019rsv}), PICO detectors at SNOLAB have reached exposures of $\sim$3~ton-days~\cite{PICO:2019vsc}, including a zero-background (observed) ton-day exposure at 3.3-keV threshold~\cite{PICO:2017tgi}.  The PICO-500 experiment, funded by the Canada Foundation for Innovation, is projected
to reach a spin-dependent WIMP-proton sensitivity at the $10^{-42}$~cm$^2$ level, still four-orders-of-magnitude \emph{above} the C$_3$F$_8$ neutrino fog.
A freon bubble chamber large enough to reach atmospheric neutrino sensitivity will require the development of a completely new detector ``inner vessel,'' i.e. the vessel containing the superheated liquid target, as no facilities exist to construct silica jars larger than those made for PICO-500, and the alpha activity of synthetic silica is too large for next generation chambers anyway.  A surface radioactivity of less than 5 nBq/cm$^2$ is required to construct a 50-ton detector sensitive to an atmospheric neutrino event within 5 years, a level of radiopurity that has been demonstrated in the large surfaces of Kamland-ZEN \cite{Gando:2020ggh}.  

\subsubsection{Dark Matter Candidates with Mass 1 - 10 GeV}
\subsubsection*{Charge and Phonon Detectors}

The Super Cryogenic Dark Matter Search (SuperCDMS) SNOLAB experiment is a current DM experiment, which is commissioning its solid-state detectors 2 km underground in the SNOLAB facility in Sudbury, Canada.  The SuperCDMS detectors are optimized to operate in either a High Voltage (HV) mode or an interleaved Z-dependent Ionization and Phonon (iZIP) mode.
The combination of these two modes allows searches at very low threshold (HV) or with excellent background rejection (iZIP)~\cite{SCDMS2017}. SuperCDMS SNOLAB expects to have sensitivity to NR masses between 0.5-5~GeV to within a decade of the neutrino fog, and to electron scattering of MeV-scale DM and/or absorption of eV-scale dark photons and/or axion-like particles (ALPs). The current generation experiment expects to be background limited by cosmogenically activated isotopes, $^3$H and $^{32}$Si, $^{210}$Pb, and dark counts. SuperCDMS has established paths to upgrade the experiment, with provisions for several scenarios, including a positive detection by another current generation experiment. The SuperCDMS collaboration is undertaking a multi-stage program to improve detector performance, as described in~\cite{supercdmsWP22}. 

\subsubsection*{Phase Change Detectors}

A second technology in the 1-10 GeV mass range is the Scintillating Bubble Chamber (SBC), using liquid-noble bubble chambers to extend the nuclear/electron recoil discrimination capability of freon-filled chambers to the sub-keV thresholds needed to reach the spin-independent neutrino fog at 1~GeV.  
Noble liquids can be superheated to a far greater degree than molecular fluids, showing sensitivity to sub-keV nuclear recoils while remaining completely insensitive to electron-recoil backgrounds~\cite{Durnford:2021cvb}.  The combination of scalability, low threshold, and background discrimination \emph{at low threshold} gives the noble-liquid bubble chamber unique capability to explore the neutrino fog in the 1--10~GeV WIMP mass range.
The Canada Foundation for Innovation has funded the construction of a 
10-kg device for SBC's first DM search~\cite{Giampa:2021wte,Alfonso-Pita:2022akn}. A 
100-eV threshold LAr experiment at the scale of PICO-500 (1-ton-year exposure) will be sufficient to explore the neutrino fog to $n=2$ at 1-GeV WIMP mass.

A pure liquid can also be placed into a meta-stable state below the freezing point. A particle interaction can then create a nucleation site, and the resulting freezing creates a signature increase in temperature and dielectric constant. This technique is similarly insensitive to electron recoil backgrounds, providing a path into the coherent neutrino scattering fog. A super-cooled liquid detector should intrinsically possess lower-energy thresholds than needed for ionization~\cite{snowball_LOI} and super-cooled water could achieve keV-scale energy thresholds. Dedicated measurements are needed to demonstrate these energy thresholds and fully characterize electron recoil discrimination capabilities of a super-cooled detector at these energies.

In the next section, we discuss experiments that are sensitive to sub-GeV DM candidates. Note that these experiments are not exclusively sensitive to sub-GeV DM and can also probe DM candidates in the 1-10 GeV mass range. 
\subsubsection{sub-GeV Dark Matter Candidates}

\subsubsection*{Noble Element Detectors}

To search for sub-GeV DM, detectors must achieve much lower energy thresholds than in the traditional mass range. For experiments with thresholds of $\sim$20 eV or higher, the noble liquid detectors described in the previous section are again an attractive technology, particularly by focusing exclusively on the ionization channel in which individual electrons can be detected with high efficiency. Achieving such thresholds comes with drawbacks, including the loss of discrimination between ERs and NRs, loss of time resolution, and a class of pathological single- and few-electron backgrounds that dominate the lowest energy bins~\cite{XENON:2021myl,DS50_2018_S2Only,LUX:2020vbj}. Still, this class of detector has the lowest demonstrated background rate per unit target mass, even in S2-only mode, allowing the most stringent DM cross-section limits to date over a wide range of DM masses and interaction models.

If current and future generations of TPCs can combine the benefits of a mature, scalable technology with the low-energy sensitivity gained through improvements described in~\cite{CF1WP2} and references therein, they will be capable of competitive sensitivity to a wide variety of existing low-mass models. The sensitivity of noble liquid detectors can also be extended by doping the target medium. At low concentrations, additives with low ionization energies can increase the scintillation and ionization yield by allowing energy that would otherwise be lost as heat to translate into additional ionization and excitation. At higher concentrations, additives with low-$A$ nuclei can serve as targets with stronger kinematic couplings to light DM particles, and additives with odd-$A$ nuclei offer sensitivity to spin-dependent interactions. Experimental efforts such as DarkSide-LowMass~\cite{GlobalArgonDarkMatter:2022kfh} and LBECA~\cite{Bernstein:2020cpc} are currently underway to develop low-threshold noble liquid TPCs optimized for the unique challenges of S2-only analyses. 

Alternatively, the NEWS-G collaboration searches for sub-GeV mass particles using spherical proportional counters (SPCs) filled with light gases, including neon, methane, and helium~\cite{NEWSG_LOI}. Similar to their liquid counterparts, SPCs must continue to lower background rates as they scale to larger mass. 

Superfluid liquid helium is also a promising medium for detection of very low energy excitations when combined with new forms of readout and will be discussed below~\cite{hertel2019helium,He4LOI,Liao:2022}. 

\subsubsection*{Charge Detectors}

Charge detectors have seen substantial improvements over the last several years, resulting in significantly reduced energy thresholds. Solid-state charge detectors, including charge-coupled devices (CCDs) and semiconductor crystals, can now efficiently detect single ionized electrons with the potential to push nuclear recoil sensitivity down to DM masses near 1~MeV, approximately an order of magnitude lower in mass than noble liquid detectors.  These detectors must contend with a similar set of background issues, most notably internal radiological backgrounds, external gamma-rays, and dark currents, and methods for minimizing these will become increasingly critical as these detectors scale in target mass.

CCDs have low noise, a small bandgap, and can be assembled within a low-background package, making them excellent targets in which to detect single charges produced by DM interactions. The DAMIC experiment has deployed an array of Si CCDs, achieving competitive sensitivity for GeV-scale WIMPs~\cite{DAMIC:2016qck}. CCD detector performance can be further improved by performing repeated, non-destructive readout of charges from a single pixel, a technology known as skipper-CCDs~\cite{Tiffenberg:2017aac}. These skipper-CCDs have significantly reduced readout noise of 0.068 $e^-$ rms/pix, allowing for precise charge counting of single electrons. The first effort utilizing these skipper-CCDs for DM searches is the Sub-Electron-Noise Skipper-CCD Experimental Instrument (SENSEI) experiment, which has reported sensitivity down to the expected $\sim$1\,MeV and whose goal is to deploy about 100-g of skipper-CCDs at SNOLAB~\cite{Crisler:2018gci,Abramoff:2019dfb,SENSEI:2020dpa}. The DArk Matter In CCDs (DAMIC-M) collaboration will also employ skipper-CCDs, aiming for a 1~kg target mass~\cite{Castello-Mor:2020jhd}. 
Looking ahead, the SENSEI and DAMIC-M collaborations have joined to propose Oscura~\cite{Aguilar-Arevalo:2022kqd}, a 10\,kg skipper-CCD experiment that will have significant physics reach for both DM with masses $\gtrsim$1\,MeV scattering off electrons and DM with masses $\gtrsim$1\,eV being absorbed by electrons.  
Due to the ability to clearly resolve single electrons, the energy detection threshold is already at its fundamental limit. Instead, the ultimate sensitivity of these experiments will be limited by overall exposure, as well as a detailed understanding of low-energy physics phenomena and detector effects. 

\subsubsection*{Phonon Detectors}

As described above, the SuperCDMS collaboration is deploying phonon detectors and they, along with similar efforts by  EDELWEISS~\cite{Arnaud:2020svb} and CRESST~\cite{Abdelhameed:2019hmk}, are moving towards even lower thresholds, based on thermal and athermal phonon readout. Most experiments are on the cusp of achieving sub-eV resolution, with detectors optimized to trigger below 20~eV. Ref.~\cite{Proceedings:2022hmu} gives a recent overview of DM searches with eV-scale energy resolutions based on R\&D detectors from CRESST, EDELWEISS, SuperCDMS, and MINER.

SuperCDMS has developed 10\,g~\cite{CPD:2020xvi} and 1\,g~\cite{Ren:2020gaq} Si detectors with 3\,eV resolution, and demonstrated 10\,eV thresholds with offline triggering algorithms~\cite{Ren:2020gaq}, based on W-TES athermal phonon sensors. Both have placed limits on DM models that extend into the eV-scale energy range~\cite{SuperCDMS:2020aus,Amaral:2020ryn} in NR and ER channels, though resolutions are still more than an order of magnitude from the 0.5~eV target for probing NR DM down to the ionization limit. Additional R\&D has demonstrated noise performance equivalent to 200~meV resolution~\cite{Fink:2020noh}, close to 1~eV thresholds. Si is well-matched to DM-electron scattering down to DM masses of 2~MeV, and is competitive for elastic NR searches with other light elements such as sapphire ($\mathrm{Al_2O_3}$) and CaWO$_4$. SuperCDMS is planning near-term tests with 1~g detectors similar to the Si HVeV with comparable baseline resolution at the eV scale~\cite{supercdmsWP22}.

CRESST, in collaboration with the $\nu$CLEUS coherent neutrino program, has achieved 19~eV thresholds in $\mathrm{Al_2O_3}$ (sapphire)~\cite{Angloher:2017sxg} and 30~eV in CaWO$_4$ (calcium tungstate)~\cite{Abdelhameed:2019hmk} crystals, respectively, and is engaged in R\&D to pursue lower thresholds with smaller detectors for next-generation particle detectors. The CRESST-III result~\cite{Abdelhameed:2019hmk} sets the best current constraints down to $\sim$200~MeV on sub-GeV DM scattering elastically off nuclei.

Semiconducting crystals read out by TES-based phonon sensors have recently demonstrated to have single electron resolution in silicon by the SuperCDMS (Si HVeV) and EDELWEISS (Ge) collaborations, showing RMS noise equivalent to 0.03~\cite{Ren:2020gaq} and 0.53~\cite{Arnaud:2020svb} electron-holes pairs, respectively. As the band gaps in Si and Ge are similar, these experiments demonstrate similar reach for DM--electron scattering and absorption processes as skipper-CCD experiments. Several experiments have near-term plans to expand to using novel substrates for phonon detection, which allows for sensitivity to {\it sub-MeV} DM candidates due to the low energy-thresholds.  SPICE, part of the TESSERACT program, is building on TES-based phonon sensor work in conjunction with SuperCDMS and MINER to expand to $\mathrm{SiO_2}$ (quartz) and sapphire substrates~\cite{SPICE}, which have good cross-section reach in comparison to other polar crystals as well as strong directionality~\cite{Mitridate:2022tnv}.  Diamond and silicon carbide are also useful as substrates for single charge detection~\cite{Kurinsky:2019pgb,Griffin:2020lgd,diamondLOI}. CRESST has demonstrated a diamond detector with 70~eV resolution~\cite{Canonica:2020omq} on a CVD substrate, and is also planning near-term R\&D to achieve sub-GeV reach in diamond. While these materials have higher band gaps, which limit the mass reach of ionization-based searches to $\geq$5\,MeV, they also ensure deeper impurity traps, making detectors made from these materials less susceptible to absorption of infrared photons, which are known to cause excess charge backgrounds. Their high dielectric strength will allow kV-scale voltage biases, which translates to charge resolution of better than 10$^{-3}$~e$^{-}$. Work on phonon-mediated charge readout as well as direct single-charge readout have been proposed. More details are discussed in ~\cite{supercdmsWP22,SPLENDOR}. A further alternative is the use of hydrogen-rich crystals coupled to TES readout to search for very low energy H-recoils from dark matter interactions~\cite{Wang:2022cyk}.

While NTDs and TES-based sensors have dominated phonon-readout for the last decade, new superconducting technologies are poised to broaden options of phonon readout, including Kinetic Inductance Detectors (KIDs) and Magnetic Microcalorimeters (MMCs). KIDs are superconducting thin-film absorbers constructed as an LC resonator in which the breaking of Cooper pairs changes their surface impedance \textemdash~an effect that can be measured and related to the initial deposition energy via an RF tone. Current energy resolutions of $<$20~eV have been demonstrated, with a roadmap down to the eV-scale~ \cite{KIDSLOI}. MMCs use thin metallic-magnetic films deposited on the surface of semiconducting or scintillating crystals for fast collection and sensing of athermal phonons with timescales on the order of microseconds, which enables   phonon-pulse shape discrimination to reduce low energy backgrounds from electron-recoils in NR searches~\cite{kim2020self}. Near-term advances to optimize for energy resolution are anticipated to achieve O(eV) energy thresholds, and tests are planned for sapphire and diamond, specifically targeting sub-GeV DM searches. 

In addition to sensors designed to measure excitations to internal bulk phonon modes, a complementary approach is to monitor the center-of-mass motion of a composite object \cite{Carney:2020xol}. Optical or electrical readout of such mechanical sensors, in either the classical or quantum regime, offers a novel approach to detecting particle dark matter on a wide range of mass scales, including both heavy \cite{Monteiro:2020wcb} and light \cite{Afek:2021vjy} candidates. These sensors are naturally directionally dependent because they monitor changes to three-momentum of the mechanical system, and can reach energy thresholds well below the eV scale \cite{Afek:2021vjy}.

A complementary approach is to employ quantum detectors being developed for Quantum Information Science (QIS), such as those based on superconducting charge qubit circuitry, as extremely low-energy (meV) phonon-sensitive threshold-detectors. Detection of dark matter would take place by way of phonons produced in a substrate breaking a Cooper pair, leading to quasiparticle tunneling into the sensors (alternatively photons in a cavity can be measured non-destructively as a shift in the resonator frequency spectrum, as for the case of dark photon/axion DM~\cite{dixit}). This presence of excess quasiparticles can be measured by multiple strategies, including monitoring changes to qubit decoherence time for weakly charge-sensitive qubits~\cite{wilen} or shifts in frequency when coupled to a readout resonator as is the case for Quantum Capacitance Detectors~\cite{echternach}. These type of sensors present an interesting possibility, especially when paired with higher threshold phonon sensors, of developing a novel, multiplexed low-threshold device.

\subsubsection*{Single Photon Detectors}

A particularly fruitful area of collaboration between HEP, astrophysics, and photonics has been in the domain of single photon counting. In the optical and NIR regime, many of the technologies discussed here are closely tied to developments aimed at survey science. This synergy has also motivated the use of other optical imaging technologies not only as secondary readout, but as the primary DM scattering target as well. The requirement for high fill factor and quantum efficiency for future UVOIR focal planes will naturally make many technologies suitable for DM detection provided the dark counts can be effectively minimized. 

One example of this synergy is the use of superconducting nanowire single-photon detectors (SNSPDs) directly as the active target to probe DM that scatters off electrons in the SNSPDs and bosonic DM that is absorbed in the SNSPDs~\cite{hochberg2019snspd,Hochberg:2021yud}. Using different materials, energy thresholds of 0.8\,eV and 0.25\,eV have been demonstrated, which would extend the mass reach of of DM detectors beyond the current $\sim$1\,MeV limits, and may be reduced further through subsequent R\&D. A benefit of this approach is that for larger energy deposits, above the eV-scale, these devices may also have sensitivity to directionality, opening a new window on background discrimination. As noted above, a key challenge to overcome as this technology matures will be fabricating them at a scale that is useful for DM experiments. Existing experiments rely on established WSi nanowire technology for DM detection at optical and near-IR wavelengths.

\subsubsection*{Scintillating Detectors}

Many crystals scintillate at optical to UV wavelengths, and a number of crystal scintillators have been proposed as DM detection targets. The majority of these have not been demonstrated with low backgrounds at the single photon level or with photon efficiency adequate to detect single scintillation events. For example, CRESST has used scintillation from calcium tungstate crystals for ER/NR discrimination, but the low light yield from events at low energy precludes the use of this detection mode below 100~eV~\cite{Angloher:2017sxg}. Thus, materials which primarily scintillate as a means of energy relaxation are needed, both to allow them to be used as primary detectors and to ensure high single photon detection efficiency correlates to thresholds at the scintillation photon energy.

Organic scintillators are promising targets for sub-GeV DM detection with electron scattering~\cite{Blanco:2019lrf,Blanco:2021hlm}. Aromatic organic compounds are naturally anisotropic due to planar carbon rings, with the lowest-energy electronic wavefunctions well-approximated by linear combinations of 2p$_z$ orbitals. The typical energy of the lowest excited state is a few eV, comparable to the 1~eV bandgap of silicon detectors. Such detectors have strong synergy with the skipper-CCD program, because a large-mass organic scintillator target can be coupled to a few CCDs with appropriate light focusing; the energy of the scintillation photon lies between the 1- and 2-electron bins, providing some protection against dark rate backgrounds.

The TESSERACT project mentioned above builds a program around coupling TES-based phonon detectors to novel detector materials, including scintillators like superfluid helium, gallium arsenide and sapphire. 
HeRALD (Helium Roton Apparatus for Light Dark Matter) is part of the TESSERACT program and proposes to instrument a vessel of superfluid $^4$He for scintillation, triplet excimer, phonon, and roton detection \cite{hertel2019helium, He4LOI}, such that the ratios of signals may be used for discrimination.  
The TESs can be used to detect scintillation photons and the quenching of triplet excimers. Alternatively, ionizing the ejected atom and then accelerating them with an electric field, the ejected ions can be calorimetrically detected by the release of the heat of adsorption, providing access to meV scale energy depositions.  

Crystal scintillators have also been identified as excellent target materials for probing DM-electron interactions, with gallium arsenide (GaAs) perhaps being the most promising~\cite{Derenzo:2016fse,Derenzo:2020hfo}. TESSERACT is developing GaAs as their primary target for electromagnetically-interacting DM due to the fact that it has a bandgap at 1.42\,eV, giving it a similar intrinsic reach to Si, and has the added benefits of intrinsic radio-purity and the capability to grow it commercially in large crystals. The phonon detectors will then be sensitive to the 1.33\,eV scintillation photons produced by interactions in the crystal, while coincident detection of phonon signals enables discrimination of electronic recoil signals from nuclear recoil and heat-only backgrounds. GaAs also seems to have little to no afterglow~\cite{Derenzo:2018plr,Vasiukov:2019hwn}. 

\subsubsection{Directional Detection}
\label{sec:directionality}
Because of our motion through a roughly
stationary galactic halo, signals of DM interacting in an Earth-bound experiment are expected to
be directional in a highly characteristic way. This smoking-gun signature
should underlie the signal of whichever DM candidate is eventually discovered -- the only problem is
that, at present, very few direct detection experiments are designed with the ability to measure it.
A large-scale recoil imaging detector not only provides an opportunity to discover direct evidence for
DM interactions, but also represents a way to conclusively confirm the nature of a detected
signal as truly that of DM~\cite{OHare:2022jnx}. A corollary to this statement is that directional detection presents the optimal
way to subtract known sources of background that could mimic such a signal such as the neutrino fog. 
 Directional detection can be achieved either by imaging the nuclear recoil trajectory directly, or by inferring the direction from a proxy variable. 

The most mature technique in this area is a low-density gas TPC, capable of direct recoil imaging, and a central goal of the CYGNUS collaboration is working towards a competitive large-scale gas TPC for a directional DM search. To realize  directional DM discovery into the neutrino fog
there are a few critical performance benchmarks that must be attained. The most important of
these are (1) angular resolution better than 30$^\circ$, (2) correct head/tail recognition at a probability
better than 75\%, (3) electron backgrounds rejected by a factor of O($10^5$), and (4) all of those limits
achieved down at sub-10-keV$_\mathrm{r}$ energies and in as high-density a gas as possible. These goals appear
feasible already down to nuclear recoil energies of 8 keV$_\mathrm{r}$ in 1 atmosphere of 755:5 He:SF$_6$~\cite{Vahsen2020}, by exploiting negative ion
drift, and limiting the maximum drift length in a 10 m$^3$ CYGNUS TPC module to around 50 cm. In the short term, CYGNUS is building 1 m$^3$ prototypes in order to optimize the configuration of such a 10 m$^3$ module. Scaling up such an experiment to the 1000 $m^3$ scale needed to probe into the neutrino fog is a daunting task, but could be possible with several common modules inside common shielding. Lastly, as described in ~\cite{OHare2022}, recoil imaging crosscuts through multiple frontiers in HEP. 

Other approaches to directional detection include anisotropic light yield in crystals, and anisotropic ionization detection due to columnar recombination in liquid or gas.
These have been demonstrated at energies higher than those relevant for DM searches, but new approaches involving magnetic fields and liquid above the normal boiling point are being investigated to reach dark matter recoil energies~\cite{Martoff}.

\section{Summary and Conclusions}

Particle DM is well-motivated and encompasses a wealth of scenarios that can explain the observed properties of DM, with characteristic mass scales from 1 eV to the Planck scale. Cosmic Frontier searches for particle DM uniquely allow the observation of DM in its astrophysical environment, ensuring that detection of a new particle can be connected to the DM puzzle. We have outlined an ensemble of upcoming and proposed direct and indirect searches for DM that will significantly improve our sensitivity to DM signals across a broad range of scenarios. Advances in calibration, modeling, and theoretical understanding of signals, backgrounds, and uncertainties, will be essential to take advantage of such improvements and permit a convincing discovery. A diverse and multi-scale experimental program, combined with support for complementary measurements, analyses, and R\&D, offers the clearest path toward a convincing detection of particle DM.

\appendix
\newpage 
\section{Tables of active, planned and concept Direct Detection experiments}
\begin{tabular}{ |p{2.5cm}||p{2cm}|p{1.5cm}|p{1.5cm}|p{2cm}|p{1.5cm}|p{1.5cm}| }
 \hline
  Name & Technology & Target & Active Mass & Experiment Location & Start Ops & End Ops \\ \hline
  \hline \hline
  \multicolumn{7}{|l|}{\textbf{Currently Running or Under Construction}} \\
  \hline \hline
  LZ & TPC & LXe & 7,000 kg & SURF & 2021 & 2026 \\ \hline	
  PandaX-4T &	TPC & LXe & 4,000 kg & CJPL & 2021 & 2025 \\ \hline
  XENONnT &	TPC & LXe &	7,000 kg & LGNS & 2021 & 2025  \\ \hline \hline
  
  DEAP-3600 & Scintillator & LAr & 3,300 kg	& SNOLAB & 2016 & 2025 \\ \hline
  Darkside-20k & TPC & LAr & 50 t & LNGS & 2027 & 2035 \\ \hline
  DAMA/LIBRA & Scintillator & NaI & 250 kg & LNGS &	2003 &  \\ \hline	
  ANAIS-112	& Scintillator & NaI & 112 kg & Canfranc & 2017 & 2022	\\ \hline
  SABRE PoP	& Scintillator & NaI & 5 kg & LNGS & 2021 & 2022	\\ \hline
  COSINE-200 & Scintillator & NaI & 200 kg & YangYang	 & 2022 & 2025 \\ \hline \hline
  
  CDEX-10 & Ionization (77K) & Ge & 10 kg & CJPL & 2016 & 	\\ \hline \hline
  
  EDELWEISS III (High Field) & Cryo Ionization / HV & Ge & 33 g & LSM & 2019 & \\ \hline
  SuperCDMS CUTE & Cryo Ionization / HV	 & Ge/Si & 5 kg/1 kg & SNOLAB &	2020 & 2022	\\ \hline					SuperCDMS SNOLAB & Cryo Ionization / HV	& Ge/Si & 11 kg/3 kg & SNOLAB &	2023 & 2028	\\ \hline \hline
  
  CRESST-III (HW Tests) & Bolometer Scintillation & CaWO4 & & LNGS & 2020 &	\\ \hline \hline	
  
  PICO-40 & Bubble Chamber & C3F8 & 35 kg & SNOLAB & 2020  & \\ \hline \hline																																		NEWS-G & Gas Drift & CH4 & & SNOLAB & 2020 & 2025	\\ \hline \hline	
  
  DAMIC-M prototype & CCD Skipper & Si & 18 g & LSM & 2022 & 2023 \\ \hline
  DAMIC-M & CCD Skipper & Si & 1 kg & LSM & 2024 & 2025 \\ \hline
  SENSEI & CCD Skipper & Si & 2 g  & Fermilab & 2019 & 2020 \\ \hline										
  SENSEI & CCD Skipper & Si & 100 g  & SNOLAB & 2021 & 2023 	\\ \hline \hline																						
   \end{tabular}

\begin{tabular}{ |p{2.5cm}||p{2cm}|p{1.5cm}|p{1.5cm}|p{2cm}|p{1.5cm}|p{1.5cm}| }
\hline
  Name & Technology & Target & Active Mass & Experiment Location & Start Ops & End Ops \\ \hline
  \multicolumn{7}{|l|}{\textbf{Planned}} \\
 \hline \hline
  
  SABRE (North) & Scintillator & NaI & 50 kg & LNGS & 2022 & 2027 \\ \hline
  SABRE (South) & Scintillator & NaI & 50 kg & SUPL & 2022 & 2027 \\ \hline					
  COSINE-200 South Pole & Scintillator & NaI & 200 kg & South Pole & 2023 &	 \\ \hline	\hline					COSINUS & Bolometer Scintillator & NaI & & LNGS & 2023 &  \\ \hline
											
  Darwin / XLZD (US LXe G3) & TPC & LXe & 50,000 kg & undetermined & 2028 & 2033 \\ \hline								ARGO & TPC or Scintillator & LAr & 300 t & SNOLAB & 2030 & 2035	\\ \hline					 \hline												CDEX-100 / 1T &	Ionization (77K) & Ge & 100-1000 kg & CJPL & 202X & \\ \hline \hline						PICO-500 & Bubble Chamber & C3F8 & 430 kg & SNOLAB & 2021 & \\ \hline \hline																																													
  
  \hline \hline
  \multicolumn{7}{|l|}{\textbf{Concept or R\&D}} \\
  \hline \hline
 Oscura & CCD Skipper & Si & 10 kg Si & SNOLAB & 2025 & 2028 \\ \hline
SBC & Bubble Chamber & LAr & 1 t & SNOLAB &  2028 & ~ \\ \hline
  SNOWBALL & Supercooled Liquid	H2O &  & & & & \\ \hline											
   DarkSide-LowMass & TPC & LAr & 1.5 t &  &  & ~ \\ \hline
  ALETHEIA & TPC & He &	&China Inst. At. Energy& & \\ \hline													TESSERACT & Cryo TES	& LHe, SiO$_2$, Al$_2$O$_3$, GaAs & & undetermined & 2026 &  \\ \hline							   
  CYGNO & Gas Directional & He + CF$_4$ & 0.5 - 1~kg & LNGS  & 2024 & ~ \\ \hline
CYGNUS & Gas Directional & He + SF$_6$/CF$_4$ & ~ & Multiple sites  & ~ & ~ \\ \hline								Windchime & Accelerometer array & & ~ & Multiple sites  & ~ & ~ \\ \hline								MAGNETO-$\chi$ & Cryogenic MMC & Diamond, Sapphire, etc. & ~ & ~ & ~ & ~ \\ \hline		
 
 \end{tabular}
 \label{table:DD}

\printbibliography
\end{document}